\documentclass[aip,jmp]{revtex4-1}

\usepackage{bbm,mathrsfs}
\usepackage{graphicx}
\usepackage{amsfonts, amsmath, amssymb}
\usepackage{amsthm}
\usepackage{psfrag,color}

\definecolor{myurlcolor}{rgb}{0,0,0.7}
\definecolor{myrefcolor}{rgb}{0.9,0,0}
\usepackage{hyperref}
\hypersetup{colorlinks, linkcolor=myrefcolor,
citecolor=myurlcolor, urlcolor=myurlcolor}

\def\textbf#1{{\bf #1}}
\def\be{\begin{equation}}
\def\ee{\end{equation}}
\def\ben{\begin{eqnarray}}
\def\een{\end{eqnarray}}
\def\eea{\end{array}}
\def\bea{\begin{array}}

\newcommand{\bei}{\begin{itemize}}
\newcommand{\eei}{\end{itemize}}

\newtheorem{thm}{Theorem}

\newtheorem{lemma}{Lemma}
\newtheorem*{lemma*}{Lemma}
\theoremstyle{definition}

\def\t{^{\mbox{\tiny T}}}
\def\bfa{{\bf a}}
\def\bfb{{\bf b}}
\def\bfc{{\bf c}}
\def\bfe{{\bf e}}
\def\bfx{{\bf x}}
\def\bfy{{\bf y}}
\def\bfv{{\bf v}}
\def\bfu{{\bf u}}
\def\e{{\rm e}}
\def\I{{\rm i}}
\def\mfg{\mathfrak{g}}
\def\mfh{\mathfrak{h}}
\def\mfl{\mathfrak{l}}
\def\h{{\cal H}}
\def\g{{\cal G}}
\def\s{{\cal S}}
\def\L{{\cal L}}
\def\tr{{\rm tr}}
\def\zm{{\bf 0}}
\def\id{{\bf 1}}
\def\reals{\mathbb{R}}

\def\beq{\begin{equation}}
\def\eeq{\end{equation}}
\newcommand{\eq}[1]{(\ref{#1})}

\begin{document}

\title{Entanglement and the three-dimensionality of the Bloch ball}

\author{Ll. Masanes}
\email{ll.masanes@gmail.com}

\affiliation{Department of Physics and Astronomy, University College London, Gower Street, London WC1E 6BT, UK}

\author{M. P.\ M\"uller}

\affiliation{Institut f\"ur Theoretische Physik, Universit\"at Heidelberg, Philosophenweg 19, D-69120 Heidelberg, Germany}

\author{D. P\'erez-Garc\'ia}

\affiliation{Dpto.~Analisis Matematico and IMI, Universidad Complutense de Madrid, 28040 Madrid, Spain}

\author{R. Augusiak}
\affiliation{ICFO-Institut de Ciencies Fotoniques, 08860 Castelldefels, Barcelona, Spain}

\begin{abstract}
We consider a very natural generalization of quantum theory by letting the dimension of the Bloch ball be not necessarily three. We analyze bipartite state spaces where each of the components has a $d$-dimensional Euclidean ball as state space. In addition to this we impose two very natural assumptions: the continuity and reversibility of dynamics, and the possibility of characterizing bipartite states by local measurements. We classify all these bipartite state spaces and prove that, except for the quantum two-qubit state space, none of them contains entangled states. Equivalently, in any of these non-quantum theories interacting dynamics is impossible.
This result reveals that ``existence of entanglement" is the requirement with minimal logical content which singles out quantum theory from our family of theories.
\end{abstract}

\pacs{
}

\pagestyle{plain}

\maketitle
\tableofcontents

\section{Introduction}\label{intro}

What consistent and physically plausible modifications of quantum theory (QT) are possible? This is a question that has come up in several fields of physics, most notably in constructions of experimental tests of QT, in quantum gravity, and in the study of correlations in quantum information theory.
Some well-known modifications of QT which are based on straightforward alterations of its mathematical formalism~\cite{Gleason, Weinberg} lead to inconsistencies~\cite{Gisin}. But here we use a different method to obtain modifications and generalizations of QT which always provides consistent theories: we choose some desirable physical features of QT and classify all theories which satisfy them. This is a double-win project: if only QT satisfies the requirements then we obtain a new axiomatization in terms of simple physical properties; on the contrary, if other theories also satisfy the requirements, then we obtain consistent alternative theories which still keep the physical features that we have chosen.
The search for alternative axiomatizations of quantum theory (QT) is an old topic that goes back to Birkhoff and von Neumann~\cite{BvN, Mackey, AlfsenShultz}. But we embrace a more operational and less mathematical approach,
initiated by Hardy's work~\cite{5RA} and continued in~\cite{Daki, MM, l1, CDP}.

In our journey beyond QT we do not want to go excessively far in theory space, so we keep within the framework of generalized probability theory~\cite{5RA, MM, l1, CDP, purification, Barnum, Wilce, Barrett}, which is based on operational  notions. For instance, the state of a system can be represented by the probabilities of some pre-established measurement outcomes, which suffices to predict the probabilities for all measurements performable to that system. On top of this foundation, the quantum features that we want to preserve are: {\em Continuous Reversibility} (for every pair of pure states there is a continuous reversible transformation which maps one state onto the other) and {\em Tomographic Locality} (the state of a composite system is characterized by the statistics of measurements on the individual components). These two axioms were introduced in~\cite{5RA} and also considered in~\cite{Daki,MM}. One of the motivations to assume the reversibility and continuity of time-dynamics is that the most fundamental theories that we know---classical or quantum---enjoy it. The axiom of Tomographic Locality has a well-defined operational meaning, but additionally, it is mathematically very natural, since it endows state-spaces of multipartite systems with the familiar tensor-product structure.

In this work we classify all continuously-reversible and locally-tomographic theories for bipartite systems where each subsystem has a state space with the geometry of the Euclidean ball (like the Bloch ball of a quantum binary system but with its dimension not being necessarily equal to three). It turns out that in all such theories, with the exception of QT, binary systems do not interact, hence, they cannot be entangled nor violate any Bell inequality. These findings push forward the results obtained in~\cite{boxworld}: the toy theory called \lq\lq{}box world\rq\rq{}, which violates all Bell inequalities maximally, does not admit any entangling reversible dynamics.

The requirement that a state space has the geometry of the Euclidean ball does not look, at first sight, physically motivated. However, some axiomatizations of QT derive this fact from physical principles as an initial step~\cite{5RA, Daki, MM, l1}, before obtaining the full structure of QT. Also, this fact is a consequence of each of the following proposed principles individually: Information Causality~\cite{IC} (see~\cite{DQT}), Branch Locality~\cite{BLP}, and ``no information gain implies no disturbance"~\cite{Pfister}. Therefore, we see the results in this paper as a kind of module which can be used in many derivations of quantum theory: as soon as a physical principle implies that the state space of a system is a Euclidean ball, one can supplement this with Tomographic Locality and Continuous Reversibility, and use our results to get most of the structure of QT (see~\cite{DQT, 3D}).

Let us finally stress that, on the mathematical side, we refine the classification of groups that act continuously and transitively on the unit sphere~\cite{Montgomery, Borel},
by providing explicit characterizations of all inequivalent linear actions on the unit sphere that are transitive.
Also, due to the Reversibility Axiom, the set of pure states is a compact homogenous space~\cite{matrix_groups}, and the full state space is an orbitope~\cite{orbitope}. As such, our work provides new ways to look at these mathematical objects, and provokes new questions within the theories of homogenous spaces and orbitopes  (Section~\ref{conc}).

The derivation of the three-dimensionality of the Bloch ball by imposing Tomographic Locality and Continuous Reversibility on the bipartite state space, has also been achieved in~\cite{Daki, MM}.
There, however, some additional strong assumptions such as the ``Subspace Axiom" were made for this purpose.
In contrast, it follows from our work that, the existence of interacting dynamics (or equivalently, the existence of entanglement or the violation of Bell inequalities) is the requirement with minimal logical content that has to be imposed in order to single out dimension three. This result has been used in~\cite{DQT} to derive the full Hilbert space formalism of QT from postulates having direct physical meaning. Also, it has been used in~\cite{3D} to derive the three-dimensionality of physical space from some operational assumptions.


\section{Results: no interaction beyond QT}\label{noint}

In this section we explain the results without introducing the framework of generalized probability theory, which is left for Section~\ref{axiomatization}. In this work we only consider bipartite systems where each constituent is a binary system. A binary system contains two perfectly distinguishable states and no more, hence, it is the generalization of a quantum two-level system or qubit.

\subsection{Two binary systems in QT}

States of two-qubit systems are represented by $4\times 4$ Hermitian matrices $\rho$ that are positive, $\rho\geq 0$, and have unit trace $\tr \rho =1$. These can be written in the following basis
\[
	\sigma_1 = \left[ \begin{array}{cc}
		0 & 1 \\
		1 & 0
	\end{array}\right], \quad
	\sigma_2 = \left[ \begin{array}{cc}
		0 & -\I \\
		\I & 0
	\end{array}\right], \quad
	\sigma_3 = \left[ \begin{array}{cc}
		1 & 0 \\
		0 & -1
	\end{array}\right],  \quad
	\id = \left[ \begin{array}{cc}
		1 & 0 \\
		0 & 1
	\end{array}\right],
\]
that is
\begin{equation}\label{rho}
	\rho = \frac{1}{4} \left( \id\otimes \id + \sum_i b_i\, \id \otimes \sigma_i +  \sum_i a_i\, \sigma_i \otimes \id +  \sum_{i,j} c_{ij}\, \sigma_i \otimes \sigma_j \right).
\end{equation}
Hence, a two-qubit state $\rho$ is specified by the three vectors $\bfa= (a_1, a_2, a_3) \in \reals^3$, $\bfb\in \reals^3$ and $\bfc \in \reals^3 \otimes \reals^3$. The condition $\rho\geq 0$ translates to some algebraic constraints for $\bfb, \bfa, \bfc$. The reduced states for each of the qubits represented by \eq{rho} are given by the partial traces
\[
	\tr_2 \rho = \frac{1}{2} \left( \id + \sum_i a_i\, \sigma_i \right), \quad
	\tr_1 \rho = \frac{1}{2} \left( \id + \sum_i b_i\, \sigma_i \right).
\]
The reduced states are characterized by the Bloch vectors $\bfa, \bfb$, which satisfy $|\bfa|, |\bfb| \leq 1$, where $|\bfa| = \sqrt{\bfa \cdot \bfa} = \sqrt{\sum_i a_i^2}$ is the Euclidean norm. Local reversible transformations act on the state as
\begin{equation}\label{rev loc trans}
	\left[ \begin{array}{c}
		\bfb \\ \bfa \\ \bfc
	\end{array}\right] \rightarrow
	\left[ \begin{array}{ccc}
		B & \zm & \zm \\
		\zm & A & \zm \\
		\zm & \zm & A \otimes B
	\end{array} \right]
	\left[ \begin{array}{c}
		\bfb \\ \bfa \\ \bfc
	\end{array}\right] ,
\end{equation}
where $A,B \in \mbox{SO}(3)$, since $\mbox{SO}(3)$ is the adjoint action of $\mbox{SU}(2)$. The matrix group corresponding to all reversible transformations of two qubits (local and non-local) is the adjoint action of $\mbox{SU}(4)$, denoted by $\g$. (See \cite{GdlT} for a characterization.)

Product states are the ones such that $\bfc = \bfa \otimes \bfb$, and as it is shown below, have no correlations. It is convenient to write product states and local transformations \eq{rev loc trans} in tensor-product form; this can be done by adopting the hat notation:
\begin{equation}\label{hat rep}
	\hat \bfa = \left[ \begin{array}{c}
		1 \\  \bfa
	\end{array} \right], \quad
	\hat A= \left[ \begin{array}{cc}
		1 & \zm \\
		\zm & A
	\end{array}\right],
\end{equation}
where $\zm$ denotes the zero matrix (or vector) with dimensions specified by the context. In accordance with the claimed form of product states and local transformations, we get
\begin{equation}\label{hat rep 2}
	\hat \bfa \otimes \hat \bfb =
	\left[ \begin{array}{c}
		1 \\  \bfb \\ \bfa \\ \bfa \otimes \bfb
	\end{array} \right],
\quad
	\hat A \otimes \hat B =
	\left[ \begin{array}{cccc}
		1 & \zm & \zm & \zm\\
		\zm & B & \zm & \zm\\
		\zm & \zm & A & \zm\\
		\zm & \zm & \zm & A \otimes B\\
	\end{array}\right].
\end{equation}
Note that here, the ordering of the components under the $\otimes$-action is not the standard one. The redundant ``1" in \eq{hat rep} and \eq{hat rep 2} is equivalent to the redundant information that $\rho$ contains, since it obeys $\tr \rho =1$. A single-qubit projective measurement is characterized by a unit-length Bloch vector $\bfx$; the probabilities for the two corresponding outcomes ``$+$'' and ``$-$'' when the state is $\bfa$ are
\[
   p(+) = \frac{1+\bfx \cdot \bfa} 2 = \frac {\hat\bfx} 2 \cdot \hat\bfa\ , \qquad
   p(-) = \frac{1-\bfx \cdot \bfa} 2 = \frac {{(-\bfx)}^\wedge} 2 \cdot \hat\bfa \ .
\]
For two-qubit systems, the joint probability of the local measurement outcomes $\bfx, \bfy$ is
\begin{equation}\label{joint prob}
	p(\bfx, \bfy) =  \frac{\hat\bfx}{2} \otimes \frac{\hat\bfy}{2} \cdot
	\left[ \begin{array}{c} 1 \\  \bfb \\  \bfa \\  \bfc \end{array} \right].
\end{equation}
As mentioned above, product states give product distributions, and hence no correlations.

\subsection{Two binary systems beyond QT}\label{bQT}

In this section we define a family of theories for bipartite systems, which is a natural generalization of the above representation for two-qubit systems. In section~\ref{axiomatization}, this family is axiomatized. These theories have arbitrary Bloch dimension $d= 2, 3, 4, \ldots$ but still satisfy equations (\ref{rev loc trans})-(\ref{joint prob}) with $\bfa, \bfb, \bfx, \bfy \in \reals^d$, $|\bfa|, |\bfb| \leq 1$, $|\bfx|, |\bfy| = 1$, $\bfc \in \reals^d \otimes \reals^d$ and $A,B \in \mbox{SO}(d)$ (below we also consider groups of reversible transformations which are proper subgroups of $\mbox{SO}(d)$). However, what is not immediately clear, is how the set of non-product states and the set of non-local reversible transformations generalize. We address these two issues in the following paragraphs.

One of the definitorial properties of these theories is that the state space of a subsystem has the geometry of a Euclidean ball, that is, states can be represented by Bloch vectors of arbitrary dimension ($\bfa \in \reals^d$ with $|\bfa| \leq 1$). One can see that these state spaces have two perfectly-distinguishable states and no more (as with qubits)---so we refer to them as binary systems. Joint states for two binary systems are represented by vectors
\begin{equation}\label{generic state}
	\left[ \begin{array}{c} 1 \\  \bfb \\  \bfa \\  \bfc \end{array} \right] \in 			
	\reals^{1+d}\otimes\reals^{1+d}.
\end{equation}
In analogy with the quantum case~(\ref{joint prob}), it is natural to assume that the joint states satisfy
\begin{equation}\label{c1}
	\frac{\hat\bfx}{2} \otimes \frac{\hat\bfy}{2} \cdot
	\left[ \begin{array}{c} 1 \\  \bfb \\  \bfa \\  \bfc \end{array} \right]
	\in [0,1]\ ,
\end{equation}
for any Bloch vectors $\bfx, \bfy$. In Section~\ref{Axioms} this condition is derived from some axioms.

Joint reversible transformations for two binary system are $(d+1)^2 \times (d+1)^2$ real invertible matrices, which map a state~(\ref{generic state}) onto another state~(\ref{generic state}). Since physical transformations can be composed, they form a group, denoted by $\g$. We consider theories where the sets of states and reversible transformations have related geometries, since we impose the following axiom.
\begin{itemize}
	\item[] {\bf Continuous reversibility:} in each type of system, for every pair of pure states there is a continuous reversible transformation mapping one state onto the other.
\end{itemize}
By continuous we mean that the evolution of the state~(\ref{generic state}) is continuous in time.
Or that every reversible transformation is part of a continuous one-parameter subgroup $\{G(t)\}_{t\in\mathbb{R}}$, where $t\in\mathbb{R}$ can be interpreted as time.
This is equivalent to saying that the group of reversible transformations is connected.
This was introduced in the axiomatization for QT given in~\cite{5RA}, under the name \lq\lq{}continuity axiom\rq\rq{}. The continuity of reversible transformations is suggested by the apparent continuity of time-evolution in the physical world.

As pointed out in~\cite{5RA}, in classical probability theory, finite dimensional systems violates this axiom, since the set of reversible transformations is the group of permutations which is not connected.
In the infinite-dimensional case, this axiom is also violated if arbitrarily sharp effects are allowed. But in classical mechanics, it makes little sense to include these unphysical measurements. For instance, they allow for computing functions that are logically uncomputable. Also, an external agent performing arbitrarily sharp measurements and preparations can violate Hamilton's equations.

Continuous reversibility implies that once the group of reversible transformations $\g$ is given, the set of states~(\ref{generic state}) is fixed, since the set of pure states is $\{G (\hat\bfa \otimes \hat\bfa) :   G\in \g\}$ for any fixed $\bfa \in \reals^d$ with $|\bfa| =1$; and the set of all states (pure and mixed) is the corresponding convex hull~\cite{convex_book}. All states generated in this fashion must give consistent probabilities, as required in~\eq{c1}, hence
\begin{equation}\label{C1}
	\mbox{$\frac{1}{4} $}
	(\hat\bfa \otimes \hat\bfa) \cdot G (\hat\bfa \otimes \hat\bfa)
	\in [0,1]\ \mbox{ for all }\ G\in \g\ .
\end{equation}
Since all pure product states can be reversibly mapped to all other pure product states, this constraint is equivalent to
\begin{equation}\label{const_1.2}
	\mbox{$\frac{1}{4} $}
	(\hat\bfx \otimes \hat\bfy) \cdot G (\hat\bfa \otimes \hat\bfb)
	\in [0,1]\ \mbox{ for all }\ G\in \g \mbox{ and } |\bfa|=|\bfb|=|\bfx|=|\bfy|= 1.
\end{equation}
Continuous Reversibility applies to all types of systems, and in particular to a single binary system. The group of reversible transformations for a binary system, denoted by $\h$, comprises
$d\times d$ real matrices $H\in \h$ which map states to states ($|\bfa|\leq 1 \Rightarrow | H\bfa | \leq 1$), and map any point in the unit sphere to any other. Table~\ref{groups trans} in Section~\ref{trans groups} contains the list of all such groups, which consists of ${\rm SO}(d)$ and some of its subgroups.  Following the hat notation~\eq{hat rep}, we denote by $\hat\h$ the representation of $\h$ which acts on the $(d+1)$-dimensional vector $\hat\bfa$. The group of local transformations for two binary systems is
$\hat\h \times \hat\h = \{\hat A \otimes \hat B : A,B\in \h\} \leq \g$, where the hat notation works as in~\eq{hat rep 2}. Clearly, local transformations constitute a subgroup of general reversible transformations $\hat\h \times \hat\h \leq \g$. Except for this and~\eq{C1} the group $\g$ is totally unconstrained.

In summary, each theory from this family is characterized by:
\begin{enumerate}
	\item the dimension of a binary system $d=2,3\ldots$,
	
	\item a group of reversible transformations for a binary system $\h$ (from Table~\ref{groups trans} with the right $d$),
	
	\item a compact connected group of $(d+1)^2 \times (d+1)^2$ real matrices $\g$ satisfying~\eq{C1} and $\hat\h \times \hat\h \leq \g$.
\end{enumerate}
For every $d$ and $\h$ there is at least one such theory: the one where only local transformations are allowed $\g = \hat\h \times \hat\h$. This type of theory has no interacting dynamics, in the sense that each subsystem evolves independently of the other. In other words, the corresponding Hamiltonians (Lie algebra elements) are of the form $H_{12} = H_1 \otimes \id_2 + \id_1 \otimes H_2$. In such theories, there are no entangled states, and Bell inequalities are not violated. However, there could be other theories within our family which violate Bell inequalities, even more than QT.
The main contribution of this work establishes that this is not the case.

\begin{thm}\label{thm1}
%
%
Let $\h$ be a group from Table~\ref{trans groups} different from ${\rm SO}(3)$, let $d$ be its associated dimension, and let $\bfa \in \reals^d$ be a unit vector. All connected groups $\g$ satisfying $\hat\h \times \hat\h \leq \g$ and $\frac{1}{4} (\hat\bfa \otimes \hat\bfa) \cdot \g (\hat\bfa \otimes \hat\bfa) \subseteq [0,1]$ are subgroups of ${\rm SO}(d)^\wedge  \times {\rm SO}(d)^\wedge$.
\end{thm}
%
\noindent All such groups $\g$ correspond to dynamics with no interaction, hence, the associated theories have no entanglement. In \cite{GdlT} the following is shown.
%
%
\begin{thm}\label{thm2}
%
%
Let $d=3$ and let $\bfa \in \reals^3$ be a unit vector. All connected groups $\g$ satisfying ${\rm SO}(3)^\wedge  \times {\rm SO}(3)^\wedge \leq \g$ and $\frac{1}{4} (\hat\bfa \otimes \hat\bfa) \cdot \g (\hat\bfa \otimes \hat\bfa) \subseteq [0,1]$ are:
	\begin{enumerate}
	\item ${\rm SO}(3)^\wedge  \times {\rm SO}(3)^\wedge$,
	\item the adjoint action of ${\rm SU}(4)$,
	\item the partially-transposed adjoint action of ${\rm SU}(4)$.
	\end{enumerate}
\end{thm}
%
%
As mentioned above, the adjoint action of ${\rm SU}(4)$ corresponds to QT for a 4-level system. Partially-transposed quantum theory and (standard) quantum theory are two representations of the same theory; since there is a reversible linear map for states, transformations and effects mapping one theory onto the other. Actually, in~\cite{GdlT}, they show a generalization of Theorem \ref{thm2} for an arbitrary number of binary systems. Theorems \ref{thm1} and \ref{thm2} imply that
\begin{center}{\em
	The only theory from the family under consideration\\ which has interacting dynamics is QT.
}\end{center}

\section{Axiomatization of the family of theories}
\label{axiomatization}

In this section we axiomatize the family of theories under consideration. But before, we introduce a framework which allows to represent states, measurements and transformations independently of the theory that we are considering.

\subsection{Generalized probability theory}\label{gpt}

In classical probability theory there can always be a joint probability distribution for all random variables under consideration. In the framework of generalized probability theory~\cite{5RA, MM, l1, CDP, purification, Barnum, Wilce, Barrett, Mielnik} this is relaxed, by allowing the possibility of random variables that cannot have a joint probability distribution, or cannot be simultaneously measured (like non-commuting observables in QT).

In this framework, a state can be represented by the probabilities of some pre-established measurement outcomes $x_1, \ldots, x_K$ which are called {\em fiducial}:
\[
	\omega= \left[ \begin{array}{c}
		p(x_1) \\ \vdots \\ p(x_K)
	\end{array} \right]
	\in\ \s\ \subset\ \mathbb{R}^{K}.
\]
This list of probabilities has to be minimal and contain sufficient information to predict the probability distribution of all measurements that can be performed on the system under consideration. We include the possibility that the system is absent, indicated by the fact that a measurement gives no outcome, hence the state space contains the null vector $\zm \in\s$. One example is the set of probability distribution normalized to any value within $[0,1]$. Another example is a spin-$\frac 1 2$ particle in QT, where the fiducial probabilities can be $[p(\sigma_1 = 1), p(\sigma_2 = 1), p(\sigma_3 = 1), p(\sigma_3 = -1)]$, and the probability that the system is present is $p(\sigma_3 = 1) + p(\sigma_3 = -1)$. Note that in classical probability theory, all fiducial outcomes are simultaneously measurable, while in QT this is not the case. Also, note that the set of fiducial outcomes need not be unique, since any three linearly independent spin directions characterize the state of the spin-$\frac 1 2$ particle. The set of all allowed states $\s$ is convex \cite{convex_book}, because if $\omega_1, \omega_2 \in\s$ then one can prepare $\omega_1$ with probability $q$ and $\omega_2$ otherwise, effectively preparing the state $q\omega_1 + (1-q)\omega_2$. The number of fiducial outcomes $K$ is equal to the dimension of $\s$, otherwise one fiducial probability would be linearly related to the others, and the list not minimal. In this work we only consider finite-dimensional state spaces.

The probability of a measurement outcome $x$ when the system is in state $\omega$ is given by a function $E_x (\omega)$. Suppose the system is prepared in the mixture \mbox{$q\omega_1 +(1-q) \omega_2$}, then the relative frequency of outcome $x$ should not depend on whether the label of the actual preparation $\omega_k$ is ignored before or after the measurement, hence
\[
	E_x \big( q\omega_1 +(1-q) \omega_2 \big) =
	q E_x (\omega_1) +(1-q) E_x (\omega_2)\ .
\]
This and the fact that when there is no state there is no outcome, $E_x (\zm)=0$, imply that $E_x$ is linear. Linear functions mapping $E:\s \to [0,1]$ are called {\em effects} and can be written as a scalar product $E(\omega)= E \cdot\omega= \sum_{i=1}^K E^i p(x_i)$. One can always measure whether there is a system or not, by checking that a measurement gives one outcome. The associated effect is denoted by $U$, and the subset of normalized states $\s^1:= \{\omega \in\s : U\cdot \omega =1\}$ must satisfy the consistency contraint $\s=\{p \omega: \omega \in \s^1 \mbox{ and } p \in [0,1] \}$. The {\em pure states} are the extreme points of $\s^1$~\cite{convex_book}.

Each type of system has associated to it: a state space $\s$, a set of measurements, and a set of transformations. A transformation is a map $T: \s \to\s$ which, for the same reason as outcome probabilities, has to be linear. A transformation $T$ is reversible if its inverse $T^{-1}$ exists and belongs to the set of transformations allowed by the theory. The set of reversible transformations of a particular state space $\s$ forms a group $\h$. Motivated by the physical interpretation,
we assume that $\s$ and $\h$ are both topologically closed.

Note that the geometry of the state space depends
on the choice of fiducial outcomes. For example, the state space of a spin-$\frac 1 2$ particle in QT is a Euclidean ball when the fiducial outcomes correspond to three orthogonal spin directions; otherwise
the state space becomes an ellipsoid. However, both geometries are related by a linear transformation.

\subsection{Composite systems}

\begin{figure}
	\centering
	\includegraphics[height=3cm]{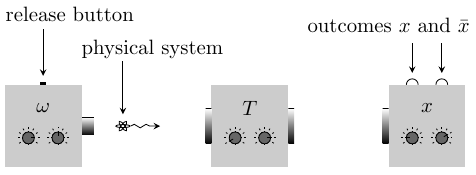}
	\caption{{\bf General experimental setup.} From left to right there are the preparation, transformation and measurement devices. As soon as the release button is pressed, the preparation device outputs a physical system in the state specified by its knobs. The next device performs the transformation specified by its knobs (which in particular can be ``do nothing"). The device on the right performs the measurement specified by its knobs, and the outcome \mbox{($x$ or $\bar{x}$)} is indicated by the corresponding light. (Reprinted from Ref. \cite{MM}.)}
\end{figure}

To a setup like Figure 1 we associate a system if, for each configuration of the preparation, transformation and measurement devices, the relative frequencies of the outcomes tend to a unique probability distribution. Two systems $A,B$ constitute a composite system $AB$ if a measurement for $A$ together with a measurement for $B$ uniquely specifies a measurement for $AB$, independently of the temporal ordering. The fact that subsystems are themselves systems implies that each has a well-defined reduced state $\omega_{A}, \omega_B$ which does not depend on which transformations and measurements are performed on the other subsystem; this is often referred to as no-signaling. Some bipartite correlations satisfying the no-signaling constraint violate Bell inequalities more than QT does~\cite{PR}; however, as we will show, these are incompatible with the axioms stated below. Naturally, system $A$ can be considered on its own or as part of a composite system $AB$, hence, for any state $\omega_A$ there is a state $\omega_{AB}$ which has $\omega_A$ as its reduced state.

A bipartite system is a system, so its states can be represented by the probabilities of some fiducial outcomes. What is the relationship between these and the fiducial outcomes of the subsystems, $x_1, \ldots, x_{K_A}$ and $y_1, \ldots, y_{K_B}$? The fact that $p(x,y)$ does not depend on the ordering of the measurements giving outcomes $x,y$ implies the following

\begin{lemma*}
%
%
The joint probability of any pair of subsystem outcomes $p(x,y)$ is given by
\begin{equation}\label{prob rule}
	p(x,y)= (E_x \otimes E_y) \cdot\omega_{AB}
\end{equation}
where
\begin{equation}\label{psiAB}
	\omega_{AB} = \left[ \begin{array}{c}
	p(x_1, y_1)\\ p(x_1, y_2)\\ \vdots\\ p(x_{K_A}, y_{K_B})
	\end{array} \right]\ \in\ \s_{AB}
	\subset \mathbb{R}^{K_A} \otimes \mathbb{R}^{K_B},
\end{equation}
and the set of all these vectors $\omega_{AB}$ spans the space $\mathbb{R}^{K_A} \otimes \mathbb{R}^{K_B}$.
\end{lemma*}

\begin{proof} If system $B$ is measured first giving outcome $y_j$, then system $A$ is in the state determined by the fiducial probabilities $p(x_i|y_j)= p(x_i, y_j)/p(y_j)$, and the single-system probability rule can be applied $p(x|y_j)= \sum_i E^i_x\, p(x_i|y_j)$. Multiplying by $p(y_j)/ p(x)$ and using the Bayes rule gives $p(y_j |x)= \sum_i E^i_x\, p(x_i,y_j) /p(x)$. By using the freedom in the ordering of measurements, we can interpret $p(y_j |x)$ as the state of system $B$ once system $A$ has been measured giving outcome $x$, and the single-system probability rule can be applied again $p(y|x)= \sum_j E^j_y\, p(y_j |x)= \sum_{i,j} E^i_x\, E^j_y\, p(x_i,y_j) /p(x)$. Multiplying both sides of this equality by $p(x)$ gives (\ref{prob rule}). Clearly, the marginal states are given by $\omega_A = (\id \otimes U) (\omega_{AB})$ and $\omega_B = (U \otimes \id) (\omega_{AB})$.

Let us see that the $\omega_{AB}$ span the full tensor product space.
In QT, the only states $\omega_{AB} \in\s_{AB}$ which have pure states as marginals $\omega_A\in \s_A , \omega_B \in \s_B$, are product ones $\omega_{AB}= \omega_A \otimes \omega_B$. The same proof technique applies to generalized probability theory. This implies that $\s_{AB}$ contains all product states, otherwise there would be a state in $\s_A$ or $\s_B$ which is not the marginal of any state in $\s_{AB}$.
Next, note that by minimality, $\s_A$ contains $K_A$ linearly independent vectors, and analogously for $\s_B$. The tensor products of these vectors are a set of $K_{AB}= K_A K_B$ linearly independent vectors in $\s_{AB}$, so the set $\s_{AB}$ has full dimension.
\end{proof}

What about global measurements? The axiom of Tomographic Locality states that the probability for the outcome of any measurement, local or global, is determined by the joint probability $p(x,y)$ of all local measurements. This implies that~(\ref{psiAB}) is a representation of a bipartite state, since all outcome probabilities can be calculated from it.

\subsection{Axioms}\label{Axioms}

\begin{figure}
   \centering
   \includegraphics[height=2.5cm]{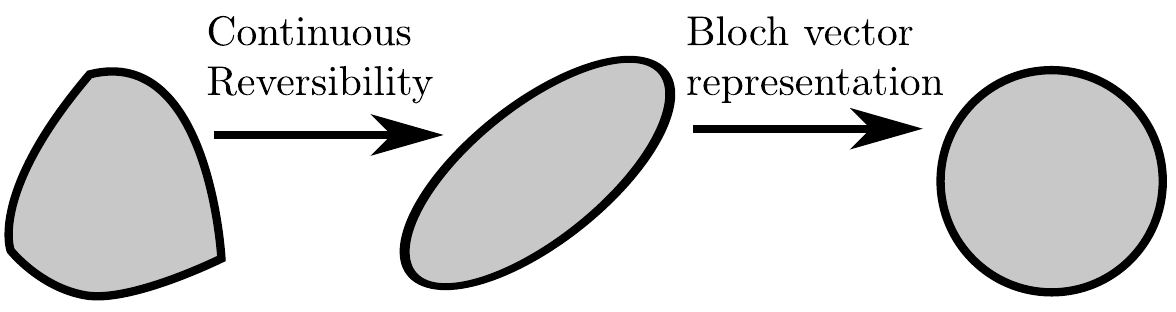}
   \caption{{\bf From Roundness to the Bloch ball.} The Roundness Axiom states that the set of normalized states of a binary system does not contain any lines in its boundary. This is true for many convex sets, like the one depicted on the left. Additionally imposing the Continuous Reversibility Axiom forces this set to be an ellipsoid. Reparametrizing the ellipsoid we end up with the Bloch ball.}
   \label{FigRoundness}
\end{figure}

The following axioms single out the family of theories defined at the beginning of Section~\ref{bQT}.
\begin{description}
	
	\item[Tomographic Locality] the state of a composite system is characterized by the statistics of measurements on the individual components.
	
	\item[Roundness] the set of normalized states of a binary system is strictly convex.

	\item[Continuous Reversibility] in each type of system, for every pair of pure states there is a continuous reversible transformation that maps one state to the other.
\end{description}
As in Section~\ref{noint}, consider a bipartite system where each subsystem is binary. Let $\s$ be the state space of a binary system, and $\s^1= \{\omega \in\s : U\!\cdot \omega =1 \}$ the subset of normalized states. The Roundness Axiom implies that the convex set $\s^1$ is \emph{strictly} convex, that is, its boundary does not contain any lines, as depicted in Fig.~\ref{FigRoundness}.
Thus, all points in the boundary represent pure states which, by the Continuous Reversibility Axiom, are reversibly connected. This enforces some additional symmetry in the state: it has to be an ellipsoid. This can be seen as follows.

Using the Haar measure on the compact group $\h$, we can define a positive matrix $W^2:=\int_{\h} H\t H\, dH$, and $W$ as its unique positive square root. For any pair of pure states $\omega, \varphi$ we have $\varphi=H\omega$ for some $H\in\h$, hence $|W\varphi|= \sqrt{\varphi\cdot W^2 \varphi}= \sqrt{\omega\cdot H\t W^2 H \omega}= \sqrt{\omega\cdot W^2 \omega}= |W\omega|$. (Note that $W\t =W$ and $W^2 H= HW^2$.) This allows to define the constant $c:=|W\omega|$ for any pure state $\omega$. The set of normalized states $\s^1$ is the intersection of the ellipsoid $\{x\in\mathbb{R}^K:  |Wx|^2\leq c^2\}$ with the normalization hyperplane $\{x\in\mathbb{R}^K : U\!\cdot x=1\}$, which is itself an ellipsoid of dimension $d=K-1$.

In what follows we reparametrize the state space to obtain the Bloch ball. Let $M:\mathbb{R}^K \to \mathbb{R}^K$ be the linear map which takes the ellipsoid $\s^1$ to the unit ball $\{\hat \bfa : |\bfa|\leq 1\}$, where
we use the hat notation~(\ref{hat rep}). This defines
a new representation for the states, measurements and transformations of a binary system as
\[
 	\omega \mapsto M \omega ,\quad
	E \mapsto E M^{-1} ,\quad
	T \mapsto M T M^{-1},\quad
      U\mapsto (1,0,\ldots,0)\t,
\]
which is the Bloch vector representation described in Section~\ref{noint}. Note that in this representation, states are no longer lists of probabilities, and all reversible transformations become orthogonal matrices which preserve the normalization. This change of representation can be applied to the state space of the bipartite system as $\omega_{AB} \to (M \otimes M) \omega_{AB}$. This gives the Bloch vector representation used in Section~\ref{noint}. Hence, the above axioms single out the family of theories defined in Section~\ref{bQT}.  For a more detailed account, see~\cite{DQT}.


\section{Proof of Theorem 1}\label{proofs}

\subsection{Groups that are transitive on the sphere}\label{trans groups}

Due to our axiom of Continuous Reversibility, the group $\h$ acting on a single binary system (a $d$-dimensional Bloch ball)
must have the following properties:
\begin{enumerate}
	\item States are mapped to states: for any $H \in \h$ and any $\bfa \in \mathbb{R}^d$ such that $|\bfa|\leq 1$ we have $| H\bfa | \leq 1$.
	\item Continuity: $\h$ is connected.
	\item Reversibility: for any $\bfa, \bfb \in \mathbb{R}^d$ such that $|\bfa| = |\bfb| = 1$ there is $H\in \h$ satisfying $\bfa = H\bfb$.
\end{enumerate}
Since the first property must hold for $H$ and its inverse $H^{-1}$, it implies that $H$ is an orthogonal matrix: $\h \leq {\rm O}(d)$. This together with connectedness implies $\h \leq {\rm SO}(d)$. The third property is called transitivity on the sphere.
Since all compact matrix groups are Lie groups~\cite{matrix_groups} we can invoke the classification in~\cite{Montgomery, Borel}: all connected compact Lie groups that act continuously, transitively and almost effectively on the sphere are the ones listed in the first column of Table~\ref{groups trans}, up to topological equivalence. However, we are interested in something more particular than continuous action: linear action. In the appendix we describe one particular representation for each of these groups, show its transitivity on the sphere, and prove its uniqueness up to linear equivalence.
We say that two matrix groups $\h, \h'$ are linearly equivalent if $\h' = \{M G M^{-1} |\, G\in\h\}$, for some invertible matrix $M$. Write $M$ in (real) polar form $M = OP$ \cite{HJ}, where $O$ is orthogonal and $P$ positive. Since $\h, \h' \leq {\rm SO}(d)$ we have $(MGM^{-1})\t (MGM^{-1}) = \id$, which implies $G P^2 = P^2 G$; and since $P$ is positive $G P = P G$, which implies $MGM^{-1} = OGO\t$. In summary: up to a change in the orthonormal coordinates of the sphere, all groups of reversible transformations for a binary system $\h$ are the ones listed in Table~\ref{groups trans} and described in the Appendix.

\begin{table}[h]\label{groups trans}
\begin{center}
\begin{tabular}{|c|c|c|}
\hline
	abstract groups & $d$ & $\h$ \\
\hline\hline
	${\rm SO}(d)$ & $3,4,5\ldots$ & ${\cal V}$ \\
\hline
	${\rm SU}(d/2)$ & $4,6,8\ldots$ & ${\cal V} \oplus {\cal V}^*$ \\
\hline
	${\rm U}(d/2)$ & $2,4,6,8\ldots$ & ${\cal V} \oplus {\cal V}^*$ \\
\hline
	${\rm Sp}(d/4)$ & $8,12,16\ldots$ & ${\cal V} \oplus {\cal V}^*$ \\
\hline
	${\rm Sp}(d/4) \times {\rm U}(1)$ & $8,12,16 \ldots$ & ${\cal V} \oplus {\cal V}^*$ \\
\hline
	${\rm Sp}(d/4) \times {\rm SU}(2)$ & $4,8,12\ldots$ & irreducible \\
\hline
	${\rm G}_2$ & $7$ & ${\cal V}$ \\
\hline
	${\rm Spin}(7)$ & $8$ & ${\cal V}$ \\
\hline
	${\rm Spin}(9)$ & $16$ & ${\cal V}$ \\
\hline
\end{tabular}
\end{center}
\caption{The first column is the list of abstract groups (or families of groups parametrized by $d$) that are transitive on the unit sphere within $\reals^d$. The second column contains the values of $d$ for which this holds. The third column schematically specifies which representation of each abstract group corresponds to the matrix group $\h$, where ${\cal V}$ is the fundamental representation and ${\cal V}^*$ its dual (both irreducible). In cases where describing the representation is complicated we just mention whether it is irreducible.}
\end{table}

Let us recapitulate the definition of some abstract groups:
\begin{eqnarray}
	{\rm SO}(n) &=& \{ Q \in \mathbb{R}^{n\times n} |\, Q\t Q =\id_n \mbox{ and } \det Q =1\}, \label{SO}\\
	{\rm SU}(n) &=& \{ Q \in \mathbb{C}^{n\times n} |\, Q^\dagger Q =\id_n \mbox{ and } \det Q =1 \}, \label{SU}\\
	{\rm U}(n) &=& \{ Q \in \mathbb{C}^{n\times n} |\,  Q^\dagger Q =\id_n \}, \label{U}\\
	{\rm Sp}(n) &=& \{ Q \in \mathbb{C}^{2n\times 2n} |\, Q^\dagger Q =\id_{2n} \mbox{ and }  Q\t JQ = J\}, \label{Sp}
\end{eqnarray}
where $J= (\I\sigma_2) \otimes \id_n$ and $\id_n$ is the $n\times n$ identity matrix. For the definition of ${\rm G}_2$ see~\cite{Arenas}, for the definition of Spin($n$) see \cite{Haar-book}. The fundamental representation ${\cal V}$ is the defining one (\ref{SO}-\ref{Sp}). According to Table \ref{groups trans}, the representation $\h$ for ${\rm SO}(d)$, denoted $\h_{{\rm SO}(d)}$, is the fundamental ${\cal V}$, hence $\h_{{\rm SO}(d)} = {\rm SO}(d)$. The representation ${\cal V} \oplus {\cal V}^*$ makes use of a standard trick to generate a real representation for a group of complex matrices. The particular map is:
\begin{eqnarray}\nonumber
	\mathbb{C}^{n\times n} &\longrightarrow& \mathbb{R}^{2n\times 2n}
	\\ \label{map}
	Q &\longmapsto& \id_2 \otimes {\rm re} Q +(\I \sigma_2) \otimes {\rm im} Q\ .
\end{eqnarray}
To see that this is a homomorphism, note that the real matrix $(\I\sigma_2)$ behaves as the imaginary unity $(\I\sigma_2)^2 =-\id_2$. This specifies the representation $\h$ for the abstract groups ${\rm SU}(d/2), {\rm U}(d/2), {\rm Sp}(d/4)$, denoted $\h_{{\rm SU}(d/2)}, \h_{{\rm U}(d/2)}, \h_{{\rm Sp}(d/4)}$. The group ${\rm SO}(d)$ with $d=2$ is not in Table \ref{groups trans} because ${\rm SO}(2) = \h_{{\rm U}(1)}$, and we choose to include it in the ${\rm U}(d/2)$ family because ${\rm SO}(2)$ is reducible, while ${\rm SO}(d)$ for $d\geq 3$ not. Another coincidence is ${\rm SU}(2) = {\rm Sp}(1)$.

The matrix group ${\cal F}_{{\rm SU}(2)}$ is the representation of ${\rm SU}(2)$ obtained through the following Lie algebra homomorphism:
\begin{eqnarray}
	\I\sigma_1 &\longmapsto& \sigma_1 \otimes (\I\sigma_2) \otimes \id_{d/4}\ , \label{s1} \\
	\I\sigma_2 &\longmapsto& (\I\sigma_2) \otimes \id_2 \otimes \id_{d/4}\ , \label{s2} \\
	\I\sigma_3 &\longmapsto& \sigma_3 \otimes (\I\sigma_2) \otimes \id_{d/4}\ . \label{s3}
\end{eqnarray}
Note that ${\cal F}_{{\rm SU}(2)}$ and ${\cal H}_{{\rm SU}(2)}$ are different representations, $\h_{{\rm SU}(2)}$ has dimension 4 and ${\cal F}_{{\rm SU}(2)}$ has dimension $d$. As shown in the Appendix, each element of ${\cal F}_{{\rm SU}(2)}$ commutes with all elements of $\h_{{\rm Sp}(d/4)}$. The $\h$-representation of ${\rm Sp}(d/4) \times {\rm SU}(2)$, denoted ${\cal H}_{{\rm Sp}(d/4) \times {\rm SU}(2)}$, is the product of all matrices from $\h_{{\rm Sp}(d/4)}$ times all matrices from ${\cal F}_{{\rm SU}(2)}$; we write this as ${\cal H}_{{\rm Sp}(d/4) \times {\rm SU}(2)} = \h_{{\rm Sp}(d/4)} {\cal F}_{{\rm SU}(2)}$. The $\h$-representation of ${\rm Sp}(d/4) \times {\rm U}(1)$ is obtained in the same way, but instead of ${\cal F}_{{\rm SU}(2)}$ we use its subgroup ${\cal F}_{{\rm U}(1)}$, with Lie algebra generated by the single element (\ref{s2}). Also note that $\h_{{\rm U}(d/2)} = \h_{{\rm SU}(d/2)}  {\cal F}_{{\rm U}(1)}$.

\subsection{Lie algebras}

Compact matrix groups are Lie groups \cite{matrix_groups}. Hence, associated to each group $\g$ there is a Lie algebra $\mfg$. If additionally, $\g$ is connected, then for each $G\in \g$ there is $X\in \mfg$ such that $G= \e^X$. All the Lie algebras that appear in this work are real vector spaces, so for any $x,y\in\reals$ and any $X,Y\in \mfg$ we have $xX+yY\in\mfg$. Another property that we use is that $GXG^{-1} \in\mfg$ for any $G\in \g$.

We denote by $\mfh, \mfg$ the Lie algebras of $\h, \g$, respectively. Since $\h \leq$SO($d$), we have $\mfh \leq \mathfrak{so}(d)$. Recall that $\mathfrak{so}(d)$ is the antisymmetric subspace of $\mathbb{R}^{d\times d}$. The group of local transformations
\beq\label{gr-loc}
	\L = \Big\{
	\left[\begin{array}{cccc}
	1 & \zm & \zm & \zm \\
	\zm & B & \zm & \zm \\
	\zm & \zm & A & \zm \\
	\zm & \zm & \zm & A\otimes B
\end{array}\right] \, |\,  A,B\in\h \Big\}
\eeq
has Lie algebra
\[
	\mfl = \Big\{
	\left[\begin{array}{cccc}
	0 & \zm & \zm & \zm \\
	\zm & Y & \zm & \zm \\
	\zm & \zm & X & \zm \\
	\zm & \zm & \zm & X\otimes\id +\id\otimes Y
\end{array}\right]\, |\, X,Y \in \mfh \Big\}
\]
where, from now on, $\id$ denotes the identity matrix of dimension specified by the context. The condition $\hat \h \times \hat \h \leq \g$ can be written as $\L\leq \g$, or equivalently $\mfl \leq \mfg$.

According to Lemma 1 from \cite{MM}, the compactness of $\g$ implies that there is $M \in \reals^{(d+1)^2 \times (d+1)^2}$ symmetric $M\t=M$ and strictly positive $M>\zm$, such that for any $G\in\g$, the matrix $M^{-1} G M$ is orthogonal. Equivalently, for any $X\in\mfg$ the matrix $M^{-1} XM$ is antisymmetric. The Lie algebra $\tilde\mfg =\{M^{-1} XM\,\,|\,\, X\in\mfg \}$
is an equivalent representation of $\mfg$ where all elements are antisymmetric. Since $\h \leq {\rm SO}(d)$, definition (\ref{gr-loc}) implies that any $G\in \L$ is orthogonal; hence $(M^{-1} GM)\t (M^{-1} GM) = \id$, which implies $M^{-2} G= G M^{-2}$, and since $M>\zm$,
\[
	GM=MG, \mbox{ for all } G \in\l\ .
\]
In cases where $\h$ is irreducible (see Table \ref{groups trans}), ${\cal L}$ is the direct sum of four irreducible representations of $\h\times\h$, as in (\ref{gr-loc}). Invoking Schur's Lemma \cite{group_book} and the positivity of $M$ we conclude that
\begin{equation}\label{M2}
	M= \left[\begin{array}{cccc}
	1 & \zm & \zm & \zm \\
	
	\zm & \beta \id & \zm & \zm \\
	
	\zm & \zm & \alpha \id & \zm \\
	
	\zm & \zm & \zm & \gamma \id
\end{array}\right], \mbox{ with } \alpha ,\beta, \gamma >0\ .
\end{equation}
It is shown in Appendix \ref{sympl} that, in the cases where $\h$ is reducible, all symmetric matrices that commute with $\h$ are proportional to the identity, hence, Schur\rq{}s Lemma implies
\begin{equation}\label{M3}
	M= \left[\begin{array}{cccc}
	1 & \zm & \zm & \zm \\
	
	\zm & \beta \id & \zm & \zm \\
	
	\zm & \zm & \alpha \id & \zm \\
	
	\zm & \zm & \zm & N
\end{array}\right], \mbox{ with } \alpha, \beta, N >0\ .
\end{equation}
The strictly-positive matrix $N$ must commute with all $(A\otimes B)$ with $A,B \in \h$.

\subsection{Block-diagonal transformations}\label{ql}

In this subsection we show that any block-diagonal reversible transformation
\begin{equation}\label{quasi loc}
\left[\begin{array}{cccc}
	1 & \zm & \zm & \zm \\
	\zm & B & \zm & \zm \\
	\zm & \zm & A & \zm \\
	\zm & \zm & \zm & C
\end{array}\right] \in \g\ ,
\end{equation}
satisfies $C= A\otimes B$ and $A,B \in {\rm SO}(d)$. This implies that block-diagonal transformations, although perhaps not implementable locally (if $A,B\not\in\h$), act independently on each subsystem. Therefore, this kind of dynamics does not let the systems interact, and does not create entanglement.

Any block-diagonal transformation (\ref{quasi loc}) can be written as $M\e^{W} M^{-1}$ where
\begin{equation}\label{quasi loc 2}
W= \left[\begin{array}{cccc}
	0 & \zm & \zm & \zm \\
	\zm & Y & \zm & \zm \\
	\zm & \zm & X & \zm \\
	\zm & \zm & \zm & Z
\end{array}\right] \in \tilde\mfg\ ,
\end{equation}
and $X,Y,Z$ are antisymmetric. Equality $C= A\otimes B$ is implied by $NZN^{-1}= X\otimes \id + \id\otimes Y$; so this is what we show next. Substituting $M\e^{\gamma W} M^{-1}$ in equation (\ref{const_1.2}) gives
\begin{equation}
	1+ \bfy \cdot \e^{\gamma Y} \bfb + \bfx \cdot \e^{\gamma X} \bfa + (\bfx\otimes \bfy) \cdot N\e^{\gamma Z} N^{-1} (\bfa \otimes \bfb) \geq 0\ .
	\label{eq33}
\end{equation}
Setting $\bfb = - \e^{-\gamma Y}\bfy$ gives $\bfx \cdot \e^{\gamma X} \bfa - (\bfx\otimes \bfy) \cdot N\e^{\gamma Z} N^{-1} (\id\otimes \e^{-\gamma Y}) (\bfa \otimes \bfy) \geq 0$, which together with the same inequality after the transformation $\bfx \rightarrow -\bfx$ implies
\begin{equation}\label{eq18}
	(\bfx\otimes \bfy) \cdot \left[ (\e^{\gamma X} \otimes\id) - N\e^{\gamma Z} N^{-1} (\id\otimes \e^{-\gamma Y})\right] (\bfa \otimes \bfy) = 0\ .
\end{equation}
Differentiating with respect to $\gamma$ at $\gamma=0$ gives
\begin{equation}\label{eq19}
	(\bfx\otimes \bfy) \cdot \left[ (X \otimes\id)- N Z N^{-1} \right] (\bfa \otimes \bfy) = 0\ ,
\end{equation}
where we have used that $\bfy\cdot Y \bfy =0$. Analogously, $\bfa=-\e^{-\gamma X} \bfx$ in~(\ref{eq33}) yields
\begin{equation}\label{eq20}
	(\bfx\otimes \bfy) \cdot \left[ (\id\otimes Y)- N Z N^{-1} \right] (\bfx \otimes \bfb) = 0\ .
\end{equation}
The space of real $d\times d$ matrices is denoted by ${\cal M}$, the subspace of symmetric ones by ${\cal M}_+$, and the subspace of antisymmetric ones by ${\cal M}_-$, so that ${\cal M} = {\cal M}_+ \oplus {\cal M}_-$. Equation (\ref{eq19}) implies that the projection onto ${\cal M} \otimes {\cal M}_+$ of $N Z N^{-1}$ is $X\otimes \id$. Equation (\ref{eq20}) implies that the projection onto ${\cal M}_+ \otimes {\cal M}$ of $N Z N^{-1}$ is $\id\otimes Y$. The combination of the two implies that $NZN^{-1} = X \otimes\id + \id \otimes Y + T$ where $T \in {\cal M}_- \otimes {\cal M}_-$. Differentiating (\ref{eq18}) two times with respect to $\gamma$ at $\gamma=0$, and using the fact that $(\bfx\otimes \bfy) \cdot \left[ T (\id\otimes Y) - (\id\otimes Y) T \right] (\bfa \otimes \bfy) = 0$ and $(\bfx\otimes \bfy) \cdot T (X\otimes\id)(\bfa \otimes \bfy) =(\bfx\otimes \bfy) \cdot  (X\otimes\id)T(\bfa \otimes \bfy)= 0$, we obtain $(\bfx\otimes \bfy) \cdot T^2 (\bfa \otimes \bfy) = 0$, which implies $\tr T^2 =0$. Since $T$ is symmetric $T^2$ is positive, hence $T=0$. Therefore $NZN^{-1} = X \otimes\id + \id \otimes Y$, which, when exponentiated, gives $C=A\otimes B$. Also, since $N$ is symmetric and $X,Y,Z$ antisymmetric we have $NZN^{-1} = N^{-1}ZN$, which together with the positivity of $N$ implies $NZ=ZN$.

\subsection{First- and second-order constraints}

For any $W\in \tilde\mfg$ the group element $M \e^{\epsilon W} M^{-1} \in\g$ must satisfy
equation~(\ref{const_1.2}). Expanding it
to the second order in $\epsilon$ we obtain
\begin{equation}\label{low1}
	0\leq \frac 1 4 \left[ \begin{array}{c} 1 \\
	\bfx \end{array} \right] \!\otimes\!
	\left[ \begin{array}{c} 1 \\
	\bfy \end{array} \right]\cdot
	M \left(\id + \epsilon W
	+\frac{\epsilon^2}{2} W^2
	+{\cal O}(\epsilon^3) \right) M^{-1}
	\left[ \begin{array}{c} 1 \\
	\bfa \end{array} \right] \!\otimes\!
	\left[ \begin{array}{c} 1 \\
	\bfb \end{array} \right]
	\leq 1\ .
\end{equation}
Now consider the special case $\bfx=\bfa$ and $\bfy=\bfb$. Then, for $\epsilon=0$, the expression~(\ref{low1}) equals unity.
Since $\epsilon$ can be positive and negative, the first order of (\ref{low1}) gives
\[
	\left[ \begin{array}{c} 1 \\
	\bfa \end{array} \right] \!\otimes\!
	\left[ \begin{array}{c} 1 \\
	\bfb \end{array} \right] \cdot
	M W M^{-1}
	\left[ \begin{array}{c} 1 \\
	\bfa \end{array} \right] \!\otimes\!
	\left[ \begin{array}{c} 1 \\
	\bfb \end{array} \right]
	= 0\ ,
\]
because otherwise, either small positive or small negative values of $\epsilon$ would yield probabilities larger than $1$.
Since the first order is zero, the constraint moves to second order
\[
	\left[ \begin{array}{c} 1 \\
	\bfa \end{array} \right] \!\otimes\!
	\left[ \begin{array}{c} 1 \\
	\bfb \end{array} \right] \cdot
	M W^2 M^{-1}
	\left[ \begin{array}{c} 1 \\
	\bfa \end{array} \right] \!\otimes\!
	\left[ \begin{array}{c} 1 \\
	\bfb \end{array} \right]
	\leq 0\ .
\]
We can get several additional inequalities by considering the lower bound in~(\ref{low1}).
For example, the special case
\[
	\frac 1 4 \left[ \begin{array}{c} 1 \\
	-\bfa \end{array} \right] \!\otimes\!
	\left[ \begin{array}{c} 1 \\
	\bfy \end{array} \right] \cdot
	M \left(\id + \epsilon W
	+\frac{\epsilon^2}{2} W^2
	+{\cal O}(\epsilon^3) \right) M^{-1}
	\left[ \begin{array}{c} 1 \\
	\bfa \end{array} \right] \!\otimes\!
	\left[ \begin{array}{c} 1 \\
	\bfb \end{array} \right]
	\geq 0
\]
equals zero for $\epsilon=0$ if $|\bfa|=1$, which implies
\begin{eqnarray*}
	\left[ \begin{array}{c} 1 \\
	-\bfa \end{array} \right] \!\otimes\!
	\left[ \begin{array}{c} 1 \\
	\bfy \end{array} \right] \cdot
	M W M^{-1}
	\left[ \begin{array}{c} 1 \\
	\bfa \end{array} \right] \!\otimes\!
	\left[ \begin{array}{c} 1 \\
	\bfb \end{array} \right]
	&=& 0\ ,
\\ 
	 \left[ \begin{array}{c} 1 \\
	-\bfa \end{array} \right] \!\otimes\!
	\left[ \begin{array}{c} 1 \\
	\bfy \end{array} \right] \cdot
	M W^2 M^{-1}
	\left[ \begin{array}{c} 1 \\
	\bfa \end{array} \right] \!\otimes\!
	\left[ \begin{array}{c} 1 \\
	\bfb \end{array} \right]
	&\geq& 0\ .
\end{eqnarray*}
By exchanging the two subsystems we get analogous constraints. In summary, the first-order equalities are
\begin{eqnarray}
\label{e++}
	\left[ \begin{array}{c} 1 \\
	\bfa \end{array} \right] \!\otimes\!
	\left[ \begin{array}{c} 1 \\
	\bfb \end{array} \right] \cdot
	M W M^{-1}
	\left[ \begin{array}{c} 1 \\
	\bfa \end{array} \right] \!\otimes\!
	\left[ \begin{array}{c} 1 \\
	\bfb \end{array} \right]
	&=& 0\ ,
\\ \label{e-+}
	\left[ \begin{array}{c} 1 \\
	-\bfa \end{array} \right] \!\otimes\!
	\left[ \begin{array}{c} 1 \\
	\bfy \end{array} \right]\cdot
	M W M^{-1}
	\left[ \begin{array}{c} 1 \\
	\bfa \end{array} \right] \!\otimes\!
	\left[ \begin{array}{c} 1 \\
	\bfb \end{array} \right]
	&=&0\ ,
\\ \label{e+-}
	\left[ \begin{array}{c} 1 \\
	\bfx \end{array} \right] \!\otimes\!
	\left[ \begin{array}{c} 1 \\
	-\bfb \end{array} \right]\cdot
	M W M^{-1}
	\left[ \begin{array}{c} 1 \\
	\bfa \end{array} \right] \!\otimes\!
	\left[ \begin{array}{c} 1 \\
	\bfb \end{array} \right]
	&=&0\ ,
\end{eqnarray}
and the second-order inequalities are
\begin{eqnarray}
\label{e++2}
	\left[ \begin{array}{c} 1 \\
	\bfa \end{array} \right] \!\otimes\!
	\left[ \begin{array}{c} 1 \\
	\bfb \end{array} \right]\cdot
	M W^2 M^{-1}
	\left[ \begin{array}{c} 1 \\
	\bfa \end{array} \right] \!\otimes\!
	\left[ \begin{array}{c} 1 \\
	\bfb \end{array} \right]
	&\leq& 0\ ,
\\ \label{e-+2}
	\left[ \begin{array}{c} 1 \\
	-\bfa \end{array} \right] \!\otimes\!
	\left[ \begin{array}{c} 1 \\
	\bfy \end{array} \right]\cdot
	M W^2 M^{-1}
	\left[ \begin{array}{c} 1 \\
	\bfa \end{array} \right] \!\otimes\!
	\left[ \begin{array}{c} 1 \\
	\bfb \end{array} \right]
	&\geq&0\ ,
\\ \label{e+-2}
	\left[ \begin{array}{c} 1 \\
	\bfx \end{array} \right] \!\otimes\!
	\left[ \begin{array}{c} 1 \\
	-\bfb \end{array} \right]\cdot
	M W^2 M^{-1}
	\left[ \begin{array}{c} 1 \\
	\bfa \end{array} \right] \!\otimes\!
	\left[ \begin{array}{c} 1 \\
	\bfb \end{array} \right]
	&\geq&0\ ,
\end{eqnarray}
for all $\bfa, \bfb, \bfx, \bfy$ with $|\bfa|= |\bfb| =|\bfx|=|\bfy|=1$.

\subsection{Imposing the first-order constraints}

The goal of this section is to show that every $W\in\tilde\mfg$ can be written in the block matrix form
\begin{equation}\label{genW}
W= \left[\begin{array}{cccc}
	0&\zm &\zm &\zm \\
	
	\zm & Y_0 & \zm & \sum_i \bfe_i\t \otimes Y_i \\
	
	\zm & \zm & X_0 & \sum_i X_i \otimes \bfe_i\t \\
	
	\zm & -\sum_i \bfe_i \otimes Y\t_i & -\sum_i X\t_i \otimes \bfe_i & \sum_j (U_j \otimes S_j + R_j \otimes V_j )
\end{array}\right].
\end{equation}
The antisymmetry of $W$ implies that all diagonal blocks (like $Y_0$ and $X_0$) are antisymmetric. Hence, the lower-right block belongs to the antisymmetric subspace of $\mathcal{M} \otimes \mathcal{M}$, that is $(\mathcal{M}_-\otimes \mathcal{M}_+) \oplus (\mathcal{M}_+ \otimes \mathcal{M}_-)$, and can be written in the Schmidt decomposition with $R_j, S_j \in \mathcal{M}_+$ and $U_j, V_j \in \mathcal{M}_-$. The two other sums (and their negative transposes) are also written in the Schmidt decomposition. In what follows we show how the zeroes in~(\ref{genW}) follow from the first-order constraints.

By adding equality~(\ref{e++}) plus equality~(\ref{e-+}) with $\bfy=\bfb$, plus equality~(\ref{e+-}) with $\bfx=\bfa$, plus equality~(\ref{e+-}) with $\bfx=-\bfa$, we obtain
\begin{equation}\label{killj}
	\left[ \begin{array}{c} 1 \\
	\zm \end{array} \right] \!\otimes\!
	\left[ \begin{array}{c} 1 \\
	\zm \end{array} \right]\cdot
	M W M^{-1}
	\left[ \begin{array}{c} 1 \\
	\bfa \end{array} \right] \!\otimes\!
	\left[ \begin{array}{c} 1 \\
	\bfb \end{array} \right]
	=0\ .
\end{equation}
By adding equality~(\ref{e++}), plus equality~(\ref{e-+}) with $\bfy=\bfb$, plus equality~(\ref{e+-}) with $\bfx=\bfa$ and $\bfb \mapsto -\bfb$, plus equality~(\ref{e+-}) with $\bfx=-\bfa$ and $\bfb \mapsto -\bfb$, we obtain
\begin{equation}\label{killZ}
	\left[ \begin{array}{c} 1 \\
	\zm \end{array} \right] \!\otimes\!
	\left[ \begin{array}{c} 1 \\
	\bfb \end{array} \right]\cdot
	M W M^{-1}
	\left[ \begin{array}{c} 1 \\
	\bfa \end{array} \right] \!\otimes\!
	\left[ \begin{array}{c} 1 \\
	\zm \end{array} \right]
	=0\ .
\end{equation}
Adding equality~(\ref{killj}) to equality~(\ref{killj}) with $\bfb \mapsto -\bfb$ yields
\begin{equation}\label{killj2}
	\left[ \begin{array}{c} 1 \\
	\zm \end{array} \right] \!\otimes\!
	\left[ \begin{array}{c} 1 \\
	\zm \end{array} \right]\cdot
	M W M^{-1}
	\left[ \begin{array}{c} 1 \\
	\bfa \end{array} \right] \!\otimes\!
	\left[ \begin{array}{c} 1 \\
	\zm \end{array} \right]
	=0\ .
\end{equation}
Analogous equations can be obtained by
permuting the two systems and exchanging the role of states and effects in equations~(\ref{killj}), (\ref{killZ}) and (\ref{killj2}). Also, by adding equation~(\ref{killj2}), plus (\ref{killj2}) with $\bfa \mapsto -\bfa$, we get
\[
	\left[ \begin{array}{c} 1 \\
	\zm \end{array} \right] \!\otimes\!
	\left[ \begin{array}{c} 1 \\
	\zm \end{array} \right]\cdot
	M W M^{-1}
	\left[ \begin{array}{c} 1 \\
	\zm \end{array} \right] \!\otimes\!
	\left[ \begin{array}{c} 1 \\
	\zm \end{array} \right]
	=0\ .
\]
These equations yield the claimed zeroes in the block matrix $W$. We can get more information if $N \propto \id$:
By adding equation~(\ref{e++}), plus equation~(\ref{e-+}) with $\bfy =\bfb$, we obtain
\[
	\left[ \begin{array}{c} 1 \\
	\zm \end{array} \right] \!\otimes\!
	\left[ \begin{array}{c} 1 \\
	\bfb \end{array} \right]\cdot
	M W M^{-1}
	\left[ \begin{array}{c} 1 \\
	\bfa \end{array} \right] \!\otimes\!
	\left[ \begin{array}{c} 1 \\
	\bfb \end{array} \right]
	=0\ .
\]
In the case where $N \propto \id$, this implies that all $Y_i$ are antisymmetric. By exchanging the roles of
the two subsystem, we obtain analogously that all $X_i$ are antisymmetric if $N \propto \id$.

\subsection{${\rm SO}(d)$ for $d\geq 4$}

In this subsection we show that, when $\h = {\rm SO}(d)$ and $d\geq 4$, the only group $\g$ satisfying our axioms is the group of local transformations $\L$. Since in this case $\h$ is irreducible, the matrix $M$ is of the form (\ref{M2}). Define the orthonormal basis
\[
	\bfe_1 = \left[ \begin{array}{c} 1\\ 0 \\ \vdots \\ 0 \end{array}\right],\
	\bfe_2 = \left[ \begin{array}{c} 0\\ 1 \\ \vdots \\ 0 \end{array}\right],
	\ \ldots\
	\bfe_d = \left[ \begin{array}{c} 0\\ 0 \\ \vdots \\ 1 \end{array}\right],
\]
the corresponding projectors $P_i = \bfe_i \bfe_i\t$, and their complements $Q_i = \id -P_i$. The stabilizer subgroup of $\h$ on the vector $\bfe_1$ is $\h_1 =\{G\in \h: G\bfe_1 =\bfe_1\} \cong$ SO($d-1$). Since the fundamental representation of SO($d-1$) is  irreducible for $d\geq 4$, Schur's Lemma states that $\int_{\h_1}\! dG\, G = P_1$ and
\[
	\int_{\h_1}\!\! dG\, G Z G^{-1}=
	z P_1 + z\rq{} Q_1\ ,
\]
for any $Z \in \reals^{d \times d}$ and some $z, z\rq{} \in \reals$ that depend on $Z$. Note that this does not hold for $d=3$, which allows quantum theory to have non-trivial dynamics and entanglement!

If $W \in \tilde\mfg$ then it is of the form~(\ref{genW}) with $X_i,Y_i$ antisymmetric, and
\begin{equation}\label{llast}
	W' = \int_{\h_1}\!\! dG\, (\hat\id \otimes\hat G) W (\hat\id \otimes\hat G)^{-1} =
\left[\begin{array}{cccc}
	0 &\zm &\zm &\zm \\
	\zm &\zm & \zm & \zm \\
	\zm &\zm & X_0 & X_1 \otimes \bfe_1\t \\
	\zm &\zm & X_1 \otimes \bfe_1 & U_1' \otimes P_1 + U_2' \otimes (\id-P_1)
\end{array}\right]
\end{equation}
is an element of $\tilde\mfg$, where $U_1', U_2'$ are antisymmetric.
The matrix $H= \id -2(P_1+P_2) \in \h$ satisfies $H \bfe_1 = -\bfe_1$ and $HP_1 H^{-1} =P_1$. Imposing constraint (\ref{e+-2}) on the element
\[
	\frac 1 2 \left( W' - (\hat\id \otimes\hat H) W' (\hat\id \otimes\hat H)^{-1} \right) =
\left[\begin{array}{cccc}
	0 &\zm &\zm &\zm \\
	\zm &\zm & \zm & \zm \\
	\zm &\zm & \zm & X_1 \otimes \bfe_1\t \\
	\zm &\zm & X_1 \otimes \bfe_1 & \zm
\end{array}\right],
\]
with $\bfb =\bfe_2$ and $\bfx =\bfa$  gives $\bfa\cdot X_1^2 \bfa \geq 0$ for all $\bfa$. The simple fact $X_1^2 = - X_1\t X_1 = -|X_1|^2 \leq \zm$ implies $X_1= \zm$. A similar argument can be made by averaging over the stabilizer subgroup in the first system (instead of the second one, as in~(\ref{llast})), obtaining $Y_1 =\zm$. Also, the same can be done with the stabilizer subgroups on the rest of vectors $\bfe_2, \ldots, \bfe_d$, obtaining $X_i = Y_i = \zm$ for all $i$. In summary: every $W\in \tilde\mfg$ must be block-diagonal (\ref{quasi loc 2}), which implies $\g \leq \L$.

\subsection{$-\id \in \h$}\label{-id}

As shown in the Appendices, the $\h$-representations of ${\rm U}(d/2)$, ${\rm Sp}(d/4)$, ${\rm Sp}(d/4) \times {\rm U}(1)$,  ${\rm Sp}(d/4) \times {\rm SU}(2)$, ${\rm Spin}(7)$ and ${\rm Spin}(9)$ contain minus the identity matrix. The group $\h_{{\rm SU}(d/2)}$ for $d$ multiple of four, also contains minus the identity. For the sake of clarity, in this subsection we use the notation $H=-\id$. If $W\in \tilde\mfg$ then it is of the form~(\ref{genW}) and
\begin{equation}\label{eq55}
	\frac 1 2 \left(W - (\hat\id \otimes\hat H) W (\hat\id \otimes\hat H)^{-1}\right) =
\left[\begin{array}{cccc}
	0 &\zm &\zm &\zm \\
	\zm &\zm & \zm & \zm \\
	\zm &\zm & \zm & \sum_i X_i \otimes \bfe_i\t \\
	\zm &\zm & -\sum_i X_i \t \otimes \bfe_i & \zm
\end{array}\right]
\end{equation}
also belongs to $\tilde\mfg$. Constraints~(\ref{e+-2}) with $\bfx =\bfa$ and (\ref{e-+2}) with $\bfy =-\bfb$ give
\begin{eqnarray*}
	-\bfa\cdot \Big[ \sum_i X_i X_i\t \Big] \bfa +
	(\bfa\otimes\bfb)\cdot \Big[ N \sum_{ij} (X_i\t X_j) \otimes (\bfe_i \bfe_j\t) N^{-1} \Big] (\bfa\otimes\bfb) \geq 0,
	\\
	\bfa\cdot \Big[ \sum_i X_i X_i\t \Big] \bfa -
	(\bfa\otimes\bfb)\cdot \Big[ N \sum_{ij} (X_i\t X_j) \otimes (\bfe_i \bfe_j\t) N^{-1} \Big] (\bfa\otimes\bfb) \geq 0 ,
\end{eqnarray*}
which together imply the equation
\[
	\bfa\cdot \Big[ \sum_i X_i X_i\t \Big] \bfa =
	(\bfa\otimes\bfb)\cdot \Big[ N \sum_{ij} (X_i\t X_j) \otimes (\bfe_i \bfe_j\t) N^{-1} \Big] (\bfa\otimes\bfb) .
\]
Summing this equation over the special cases $\bfa, \bfb \in \{\bfe_1\ldots \bfe_d\}$ gives
\[
	d \sum_i \tr\! \left[X_i X_i\t \right] =
	\sum_{ij} \tr\! \left[ N  (X_i\t X_j) \otimes (\bfe_i \bfe_j\t) N^{-1} \right]=
	\sum_{i} \tr\! \left[  X_i\t X_i \right],
\]
which is only possible if $X_1 =\cdots =X_d =\zm$. An analogous argument shows $Y_1 =\cdots =Y_d =\zm$. Therefore, all elements of $\g$ are block-diagonal and non-interacting as in (\ref{quasi loc}).

\subsection{${\rm SU}(d/2)$ for $d\geq 6$}

In this subsection we show that, when $\h = \h_{{\rm SU}(d/2)}$ and $d\geq 6$, all groups $\g$ are non-interacting. (The case $\h_{{\rm SU}(d/2)}$ with $d=4$ is analyzed in Section \ref{-id}.) The stabilizer of $\h$ on the vector $\bfe_1$ is $\h_1 =\{G\in \h: G\bfe_1 =\bfe_1\}$, or more concretely,
\[
\h_1 = \left\{ \left.
\left[\begin{array}{cccc}
	1 &\zm & 0 &\zm \\
	\zm & {\rm re} U & \zm & {\rm im} U \\
	0 &\zm & 1 &\zm \\
	\zm & -{\rm im} U & \zm & {\rm re} U
\end{array}\right]\,\, \right|  \,\,
U \in {\rm SU}(d/2-1)
\right\}.
\]
One can check that $\int_{\h_1}\! dG\, G = P_1 + P_{d/2}$ and, for any $Z\in \reals^{d \times d}$,
\begin{equation}\label{W depol}
\int_{\h_1}\!\! dG\, G \left[\begin{array}{ccc}
	z_{1,1} & \cdots & z_{1,d} \\
	\vdots &  & \vdots \\
	z_{d,1} & \cdots & z_{d,d} \\
\end{array}\right]
 G^{-1} =
\left[\begin{array}{cccc|cccc}
	z_{1,1} & & & & z_{1,d/2} & & & \\
	& z &  &  & & z' &  &  \\
	& & \ddots &  &  &  & \ddots &  \\
	& & & z & & & & z' \\
\hline
	z_{d/2,1} & & & & z_{d/2,d/2} & & & \\
	& -z' &  &  & & z &  &  \\
	& & \ddots &  &  &  & \ddots &  \\
	& & & -z' & & & & z
\end{array}\right],
\end{equation}
for some $z, z\rq{} \in \reals$ (all blank entries in the right-hand side are zeros). If $W\in \tilde\mfg$ then
\[
	W' = \int_{\h_1}\!\! dG\, (\hat\id \otimes\hat G) W (\hat\id \otimes\hat G)^{-1} =
\left[\begin{array}{cccc}
	0 &\zm &\zm &\zm \\
	\zm & Y_0 & \zm & \sum_i \bfe_i\t \otimes Y_i \\
	\zm &\zm & X_0 & X_1 \otimes \bfe_1\t + X_{d/2} \otimes \bfe_{d/2}\t\\
	\zm & \Box & \Box & \sum_j (U_j \otimes S_j + R_j \otimes V_j )
\end{array}\right]
\]
belongs to $\tilde\mfg$ too, and $Y_0, Y_i, V_j, S_j$ are of the same form as the right-hand side of (\ref{W depol}).
The antisymmetry of $W\rq{}$ makes the $\Box$s unambiguous. Define
\[
H=
\left[\begin{array}{cc}
	U & \zm \\
	\zm & U
\end{array}\right]
\mbox{ where }\
U=\left[\begin{array}{ccccc}
	-1 & & & &  \\
	& -1 &  &  & \\
	& & 1 &  & \\
	& & & \ddots &  \\
	& & & & 1
\end{array}\right],
\]
and  the blanks are zeros. Note that the right-hand side of (\ref{W depol}) commutes with $H$. The element
\[
	\frac 1 2
	\left( W'- (\hat\id \otimes\hat H) W' (\hat\id \otimes\hat H)^{-1} \right)=
\left[\begin{array}{cccc}
	0 &\zm &\zm &\zm \\
	\zm & \zm & \zm & \zm \\
	\zm &\zm & \zm & X_1 \otimes \bfe_1\t + X_{d/2} \otimes \bfe_{d/2}\t\\
	\zm & \zm & \Box & \zm
\end{array}\right] \in \tilde\mfg\ ,
\]
has the same structure as (\ref{eq55}); therefore, arguing in the same way one obtains $X_1 = X_{d/2} =\zm$. Repeating this argument with the stabilizer of the vectors $\bfe_2, \ldots, \bfe_{d/2-1}$ gives $X_i =0$ for all $i$, and analogously for $Y_i$. Therefore, all elements of $\g$ are block-diagonal and  non-interacting as in (\ref{quasi loc}).

\subsection{${\rm G}_2$}\label{G2s}

In this section we consider the case $\h = {\rm G}_2$ and show that all the corresponding groups $\g$ are non-interacting. Since $\h$ is irreducible, $M$ has the form (\ref{M2}). Schur\rq{}s Lemma \cite{group_book} together with irreducibility imply that any $W\in \tilde\mfg$, which a priori has the generic structure (\ref{genW}), satisfies
\[
	\int_{\h}\!\! dA \,
	(\hat A \otimes \hat\id) W (\hat A \otimes \hat\id)^{-1}
	=
\left[\begin{array}{cccc}
	0 &\zm &\zm &\zm \\
	\zm & Y_0 & \zm & \zm \\
	\zm &\zm & \zm & \zm \\
	\zm & \zm & \zm
	& \id\otimes V
\end{array}\right] \in \tilde\mfg\ .
\]
In addition, and according to Section~\ref{ql}, the above element must satisfy $V=Y_0$. This implies that for any element $W\in \tilde\mfg$, there is another one
\begin{eqnarray*}
	W' &=& W-
	\int_{\h}\!\! dA \,
	(\hat A \otimes \hat\id) W (\hat A \otimes \hat\id)^{-1}  -
	\int_{\h}\!\! dB \,
	(\hat \id \otimes \hat B) W (\hat\id \otimes \hat B)^{-1}
\\ &=&
   	\left[\begin{array}{cccc}
		0&\zm &\zm &\zm \\
		\zm & \zm & \zm & \sum_i \bfe_i\t \otimes Y_i \\
		\zm & \zm & \zm & \sum_i X_i \otimes \bfe_i\t \\
		\zm & -\sum_i \bfe_i \otimes Y\t_i & -\sum_i X\t_i \otimes \bfe_i &
		\sum_j (U'_j \otimes S'_j + R'_j \otimes V'_j )
	\end{array}\right] \in \tilde\mfg\ ,
\end{eqnarray*}
with identical non-diagonal blocks, and null second and third diagonal blocks. (The fourth diagonal block might get modified, but we do not care.)

The stabilizer subgroup of $\h$ on the vector $\bfe_1$ is $\h_1 =\{G\in \h: G\bfe_1 =\bfe_1\}$. It turns out that $\h_1 \cong \h_{{\rm SU}(3)}$, which is transitive on the 6-sphere. Hence, all vectors left invariant by $\h_1$ are proportional to $\bfe_1$, therefore $\int_{\h_1}\!\! dG\, G= P_1$. According to Appendix~\ref{SecG2},  for any $Z \in \reals^{7\times 7}$ we have
\[
	\int_{\h_1}\!\! dA\ A Z A^{-1}=
	z_1 P_1 + z_2 Q_1 + z_3 T \ ,
\]
where $T$ is defined in~(\ref{def T}) and $z_1, z_2, z_3 \in \reals$. The matrix $T$ commutes with $\h_1$ and satisfies: $T\t=-T$, $T\bfe_1=0$, and $T^2=-Q_1$. All the above implies that
\begin{eqnarray*}
	W\rq{}\rq{} &=& \int_{\h_1}\!\! dA \int_{\h_1}\!\! dB\,
	(\hat A \otimes \hat B) W' (\hat A \otimes \hat B)^{-1}
\\&=&
	\left[\begin{array}{cccc}
		0 &\zm &\zm &\zm \\
		\zm &\zm & \zm & y\, \bfe_1\t \otimes T \\
		\zm &\zm & \zm & x\, T\otimes \bfe_1\t \\
		\zm &y\, \bfe_1 \otimes T & x\, T\otimes \bfe_1
		& T\otimes S + R \otimes T
	\end{array} \right] \in\tilde\mfg\ , 
\end{eqnarray*}
where $x= \frac 1 6 \tr (TX_1)$, $y= \frac 1 6 \tr (TY_1)$, and $R,S$ are linear combinations of $P_1, Q_1$. The matrix $H\in \h$ defined in~(\ref{def H}) satisfies: $H\bfe_1 = -\bfe_1$, $HTH^{-1} = -T$, $HP_1 H^{-1} = P_1$, and $HQ_1 H^{-1} = Q_1$. This allows for the construction of the following element
\[
	W\rq{}\rq{}\rq{} =  \frac 1 2 \left(W\rq{}\rq{} +
	(\hat H \otimes \hat H) W\rq{}\rq{} (\hat H \otimes \hat H)^{-1} \right)
=
	\left[\begin{array}{cccc}
		0 &\zm &\zm &\zm \\
		\zm &\zm & \zm & y\, \bfe_1\t \otimes T \\
		\zm &\zm & \zm & x\, T\otimes \bfe_1\t \\
		\zm &y\, \bfe_1 \otimes T & x\, T\otimes \bfe_1 & \zm
	\end{array}\right].
\label{eq67}
\]
Imposing (\ref{e-+2}) with $\bfa =\bfb =\bfy =\bfe_2$ we obtain $x^2-y^2 \geq 0$. Imposing (\ref{e+-2}) with $\bfa =\bfb =\bfx =\bfe_2$ we obtain $-x^2 +y^2 \geq 0$. These two inequalities imply $x=\pm y$.

In what follows we show that for any $W\in \tilde\mfg$ we have $\tr (TX_1) = \tr (TY_1) =0$, or equivalently $x=y=0$. We do this by assuming the opposite:
\[
	W_\pm
=
	\left[\begin{array}{cccc}
		0 &\zm &\zm &\zm \\
		\zm &\zm & \zm & \pm \bfe_1\t \otimes T \\
		\zm &\zm & \zm & T\otimes \bfe_1\t \\
		\zm &\pm \bfe_1 \otimes T & T\otimes \bfe_1 & \zm
	\end{array}\right] \in \tilde\mfg\ ,
\]
and obtaining a contradiction. For each $A\in\mfh$ we have a local transformation
\[
	L=\left[\begin{array}{cccc}
	0 & \zm & \zm & \zm \\
	\zm & \zm & \zm & \zm \\
	\zm & \zm & A & \zm \\
	\zm & \zm & \zm & A\otimes\id
\end{array}\right]\in\mfl \leq \tilde\mfg\ .
\]
Since the Lie algebra $\mfg$ is closed under commutations we have
\[
   \left[ \strut[L, W_\pm], W_\pm\right] =\left[
      \begin{array}{cccc}
         0& \zm & \zm & \zm  \\
         \zm & \zm & \zm & \zm \\
         \zm & \zm & [[A,T],T] & \zm \\
         \zm & \zm & \zm &  (A P_1 +P_1 A) \otimes T^2 + [[A,T],T] \otimes P_1
      \end{array}
   \right] \in \tilde\mfg\ .
\]
According to Section~\ref{ql}, the above implies $(A P_1 +P_1 A) \otimes T^2 + [[A,T],T] \otimes P_1 = [[A,T],T] \otimes \id$. Since $T^2 = -Q_1 = -(\id -P_1)$, this is equivalent to $[[A,T],T] = -(A P_1 +P_1 A)$. We can see that this equality is false by substituting $A$ by a generic element from $\mfh$ of the form~(\ref{generic G2}); therefore $x=y=0$.

The above shows that any $W\in \tilde \mfg$ satisfies $\int_{\h_1} dG\, G X_1 G^{-1} = \zm$, hence
\begin{eqnarray}
	W\rq{}\rq{}\rq{}\rq{} &=&
	\int_{\h_1}\!\!  dG\, (\hat G\otimes\id) W' (\hat G\otimes\id)^{-1}
\nonumber \\ \label{e64} &=&
	\left[\begin{array}{cccc}
		0 & \zm & \zm & \zm \\
		\zm & \zm & \zm & \bfe_1\t \otimes Y_1 \\
		\zm & \zm & \zm & \zm \\
		\zm & \bfe_1\otimes Y_1 & \zm &
		P_1\otimes V+Q_1\otimes V' +T\otimes S
	\end{array}\right] \in \tilde \mfg\ .
\end{eqnarray}
If we use $H\in \h$ defined in~(\ref{def H}) we obtain
\begin{equation}\label{las}
   	\frac 1 2\left( W\rq{}\rq{}\rq{}\rq{}
   	-(\hat H\otimes\id) W\rq{}\rq{}\rq{}\rq{} (\hat H\otimes\id)^{-1}\right)
=
	\left[
      \begin{array}{cccc}
         0 & \zm & \zm & \zm \\
         \zm & \zm & \zm & \bfe_1\t \otimes Y_1 \\
         \zm & \zm & \zm & \zm \\
         \zm & \bfe_1\otimes Y_1 & \zm & T\otimes S
      \end{array}
   \right]\in \tilde\mfg\ .
\end{equation}
Imposing constraint~(\ref{e+-}) on the above element, and using the fact that $\bfb\cdot Y_1 \bfb =0$ for any $\bfb$, we obtain $S=\zm$.
Constraint~(\ref{e-+2}) with $\bfy=\bfb$ and $\bfa=\bfe_2$ on element~(\ref{las}) yields $\bfb\cdot Y_1^2 \bfb\geq 0$ for
all $\bfb$. Since $Y_1^2=-Y_1\t Y_1\leq 0$, it follows that $Y_1=\zm$.

By integrating over the stabilizer subgroup on the vector $\bfe_1$ for the first system~(\ref{e64}), we have shown that any element $W\in \tilde \mfg$ has $Y_1=\zm$. By doing the same procedure for the second system we obtain $X_1= \zm$. Analogously, by considering the stabilizers of all vectors $\bfe_i$, we obtain $X_i = Y_i = \zm$ for all $i$. Therefore, all elements in $\tilde\mfg$ are block-diagonal~(\ref{quasi loc 2}), and the group $\g$ has non-interacting dynamics.

\section{Conclusions}\label{conc}

In this work we have explored the existence of entanglement beyond quantum theory. We have classified all continuously-reversible and locally-tomographic theories for bipartite systems where each subsystem has a Euclidean ball as its state space. We have shown that the only theory within this family which has interacting dynamics is QT; and all the others do not allow for entanglement nor violation of Bell inequalities. These results illustrate how restrictive is the requirement of interacting reversible dynamics.

It remains open the possibility that non-three-dimensional Euclidean balls display some type of multipartite entanglement~\cite{DB}. Or the existence of interacting dynamics between Euclidean balls of different dimension. However, if this last thing happens, it must be in a way which does not allow for engineering an interaction between two systems of the same type through a system of a different type, since we have shown that this is not possible. We leave all these problems for future research.

More generally, we introduce a construction which is not specific to Euclidean ball subsystems nor to continuous dynamics. It can be argued that, under the sole assumptions of Tomographic Locality and (not necessarily continuous) Reversibility, the requirement that two systems can interact is very restrictive. Consider the composition of two systems, with respective groups of reversible transformations $\h_1$ and $\h_2$.
The group of global transformation $\g$ contains the local transformations $\hat \h_1 \otimes \hat \h_2$. But does it contain a transformation $G$ not being of the form $\hat A \otimes \hat B$? If this is the case, then $\g$ must also contain all the chains of products of $G$ with the elements of $\hat \h_1 \otimes \hat \h_2$. In the generic case, these chains of products do generate the whole group of orthogonal matrices $\g \cong O(d_1 +d_2 +d_1 d_2)$ (even when $\h_1$ and $\h_2$ are finite), which violates the consistency condition~(\ref{const_1.2}).
Therefore, only for very particular choices of $\h_1$, $\h_2$ and $G$, the consistency condition is going to be satisfied. Classifying these matrix groups would be an important goal. Since it would lead to a reconstruction of QT with the certified logical minimality of its axioms.

Within generalized probability theories, entanglement is a generic feature of bipartite state spaces. However, all the above suggests that, once we restrict to  theories satisfying Reversibility and Tomographic Locality, entanglement and violation of Bell inequalities are very singular phenomena.


\begin{acknowledgements}
LlM acknowledges support from CatalunyaCaixa, EU ERC Advanced Grant NLST, EU Qessence project, the Templeton Foundation and the FQXi large grant project ``Time and the structure of quantum theory".
%
%
DP-G acknowledges support from the Spanish grants I-MATH, MTM2008-01366, S2009/ESP-1594 and the European project QUEVADIS. RA acknowledges support from AQUTE and TOQATA and the Spanish MINECO for the Juan de la Cierva scholarship.
\end{acknowledgements}


\appendix

\section{Linear actions transitive on the sphere}

We are interested in matrix groups $\h$ which are subgroups of ${\rm SO}(d)$ and map any unit vector in $\reals^d$ to any other. This implies that the action of $\h$ on $\reals^d$ has no invariant subspaces, or in other words, it is $\reals$-irreducible. However, this does not imply that $\h$ is $\mathbb{C}$-irreducible. According to~\cite{group_book} two possibilities can happen:
\begin{enumerate}
\item $\h$ is not $\mathbb{C}$-irreducible. The fact that $\h$ is $\reals$-irreducible implies that it is the direct sum of a $\mathbb{C}$-irreducible representation ${\cal V}$ and its dual ${\cal V}^*$, and the map which takes ${\cal V}$ to $\h$ is (\ref{map}). As shown below, examples of this sort are $\h_{{\rm SU}(d)}, \h_{{\rm Sp}(d)}$.

\item $\h$ is $\mathbb{C}$-irreducible. The fact that $\h$ is a real representation implies $\int_\h \!\!dG\, \tr G^2 =1$. As shown below, examples of this sort are $\h_{{\rm SO}(d)}, \h_{{\rm G}_2}, \h_{{\rm Spin}(7)}, \h_{{\rm Spin}(9)}$.
\end{enumerate}

\section{The special unitary group}

\subsection{Description}

According to definition (\ref{SU}) the Lie algebra of ${\rm SU}(n)$ is
\begin{equation}\label{su}
	\mathfrak{su}(n) = \{ X\in \mathbb{C}^{2n\times 2n}|\, X^\dagger =-X \mbox{ and } \tr X=0\}\ .
\end{equation}
According to the map (\ref{map}) the Lie algebra of $\h_{{\rm SU}(d/2)}$ with $d=2n$ is
\[
	\mfh_{{\rm SU}(d/2)} = \{ \id_2 \otimes A +
	(\I \sigma_2) \otimes S\, |\,
	A \in {\cal M}_- , S \in {\cal M}_+ \mbox{ and } \tr S=0 \}\, ,
\]
where ${\cal M}_+$ and ${\cal M}_-$ are the sets of $n\times n$ symmetric and antisymmetric real matrices, respectively. One can check that all matrices that commute with $\mfh_{{\rm SU}(d/2)}$ are of the form
\begin{equation}\label{comm SU}
	z_0 \id_n+ z_1 (\I \sigma_2) \otimes \id_{n}\ ,
\end{equation}
with $z_0, z_1 \in \reals$ if we want them to be real. Hence, all symmetric matrices which commute with $\h_{{\rm SU}(d/2)}$ are proportional to the identity.

\subsection{Transitivity}

The transitivity of $\h_{{\rm SU}(d/2)}$ follows from the transitivity of ${\rm SU}(d/2)$ in the set of unit vectors in $\mathbb{C}^{d/2}$.

\subsection{Uniqueness}

In what follows we show that $\h_{{\rm SU}(d/2)}$ is the unique $d$-dimensional real representation of ${\rm SU}(d/2)$ which is $\reals$-irreducible. This is done by first, showing that there are no $d$-dimensional real representations which are $\mathbb{C}$-irreducible, and second, showing that the only $d/2$-dimensional $\mathbb{C}$-irreducible representations are the fundamental, defined in~(\ref{SU}), and its dual (which coincide for $n=2$).

\begin{lemma}
All $2n$-dimensional $\mathbb{C}$-irreducible representations of $\mathfrak{su}(n)$ are complex.
\end{lemma}
\begin{proof}
There is a one-to-one correspondence between $\mathbb{C}$-irreducible representations of $\mathfrak{su}(n)$ and $(n-1)$-dimensional vectors of natural numbers $(a_1,\ldots, a_{n-1})$, which are the coefficients of the dominant weight when it is expressed in the basis of the fundamental weights~\cite{group_book}. The dimension of the representation is given by
\begin{equation}\label{D}
   D(a_1,\ldots, a_{n-1}) = \prod_{j=1}^n \prod_{i=1}^{j-1}\left( 1 + \frac{a_i+\ldots + a_{j-1}}{j-i}\right)\ ,
\end{equation}
so the trivial representation corresponds to $D(0,\ldots ,0)=1$~\cite{group_book}. Let us start with the case $n\geq 6$. First note that
\[
   D(2,0,0,\ldots)=\prod_{j=2}^n \left(1+\frac{a_1+\ldots + a_{j-1}}{j-1}\right)=\prod_{j=2}^n \left ( 1+\frac 2 {j-1}\right)
   =\frac{n(n+1)}2\ .
\]
Similarly,
\[
   D(0,\ldots,0,2)=\prod_{i=1}^{n-1} \left( 1 + \frac{a_i+\ldots, + a_{n-1}}{n-i}\right)=\prod_{i=1}^{n-1} \left(1+\frac 2 {n-i}\right)=\frac{n(n+1)}2\ .
\]
Splitting off the $j=n$ term and then the $i=1$ term yields
\begin{eqnarray*}
   && D(1,0,0,\ldots,0,0,1)\\
   &=&\prod_{j=2}^{n-1} \prod_{i=1}^{j-1} \left( 1 + \frac{a_i + \ldots + a_{j-1}}{j-i}\right)
   \cdot \prod_{i=1}^{n-1} \left(1 + \frac{a_i + \ldots + a_{n-1}}{j-i}\right) \\
   &=& \prod_{j=2}^{n-1} \left( 1 + \frac{a_1 + \ldots + a_{j-1}}{j-i}\right) \cdot \left( 1+\frac{a_1+\ldots + a_{n-1}}{n-1}\right)
   \cdot \prod_{i=2}^{n-1}\left(1+\frac{a_i + \ldots + a_{n-1}}{n-i}\right)\\
   &=& \prod_{j=2}^{n-1} \left(1+\frac 1 {j-1}\right) \cdot\left(1+\frac 2 {n-1}\right) \cdot \prod_{i=2}^{n-1} \left( 1 + \frac 1 {n-i}\right) = n^2-1\ .
\end{eqnarray*}
Now suppose there were $a_1,a_{n-1}$ such that $D(a_1,0,0,\ldots,0,0,a_{n-1})=2n$. We know that $a_1=1,a_{n-1}=0$ and $a_1=0,a_{n-1}=1$ are
impossible (they yield $n$ instead of $2n$). Now suppose that $a_1\neq 0$ and $a_{n-1}\neq 0$, then by strict monotonicity of $D$, we get
$D(a_1,0,\ldots,0,a_{n-1})\geq D(1,0,\ldots,0,1)=n^2-1 > 2n$.

Thus, one of $a_1$ and $a_{n-1}$ must be zero, if we want that $D$ attains the value $2n$. Suppose that $a_{n-1}=0$, then we must have $a_1\geq 2$, so
$D(a_1,0,\ldots,0)\geq D(2,0,0,\ldots)=n(n+1)/2>2n$. The same conclusion holds the other way round. In summary,
$D(a_1,0,0,\ldots,0,0,a_{n-1})=2n$ has no solutions. Therefore, for every representation of dimension $2n$, there must be at least one $j\in[2,n-2]$ such that $a_j\geq 1$. But then,
\[
	D(a_1,\ldots,a_{n-1}) \geq
	D(0,\ldots,0,\underbrace{1}_j,0,\ldots,0) =
	{n\choose j}\geq {n\choose 2} =
	n(n-1)/2>2n
\]
if $n\geq 6$. Thus, no such representation exists.

Due to strict monotonicity of $D$, we can check by hand the cases $n=3,4,5$, since only finitely many values have to be checked numerically.
It turns out that the only irreps of dimension $2n$ are:
\begin{itemize}
\item[] $n=3$: $(0,2)$ and $(2,0)$.
\item[] $n=4$: none.
\item[] $n=5$: $(0,1,0,0)$ and $(0,0,1,0)$.
\end{itemize}
According to Prop.\ 26.24 of~\cite{group_book}, these representations are all complex. This proves the lemma.
\end{proof}

\begin{lemma}
The only $n$-dimensional $\mathbb{C}$-irreducible representations of $\mathfrak{su}(n)$ are the fundamental~(\ref{su}) and its dual (which coincide for $n=2$).
\end{lemma}
\begin{proof}
The case $n=2$ is very well known, so let us assume $n\geq 3$. Since all factors in~(\ref{D}) are stricly positive, we observe that $D(a_1,\ldots, a_{n-1})$ is strictly increasing in every $a_k$. One can calculate that
\[
	D(0,\ldots, 0, \underbrace{1}_j , 0,\ldots, 0)={n \choose j}\ .
\]
In particular
\[
   D(1,0,\ldots,0)=n\ ,\qquad D(0,\ldots,0,1)=n\ ,\qquad D(0,\ldots, 0, \underbrace{1}_{j\not\in\{1,n-1\}} , 0,\ldots, 0) > n\ .
\]
Since $D$ is strictly increasing in every entry, $(1,0,\ldots,0)$ and $(0,\ldots,0,1)$ are the only arguments where $D$ attains the value $n$.
Since these two representations satisfy $a_1\neq a_{n-1}$, they must be complex according to Prop.\ 26.24 of~\cite{group_book}.
\end{proof}

\section{The symplectic group}\label{sympl}

\subsection{Description}

According to definition (\ref{Sp}), the Lie algebra of ${\rm Sp}(n)$ is
\begin{equation}\label{sp}
	\mathfrak{sp}(n) = \{ X\in \mathbb{C}^{2n\times 2n} |\, X^\dagger +X = \zm \mbox{ and } X\t J + J X = \zm\}\ .
\end{equation}
Using a representation with $n\times n$ real matrix blocks
\[
	J= \left[ \begin{array}{cc} \zm & \id_n \\ -\id_n & \zm \end{array} \right]
	, \quad
	X= \left[ \begin{array}{cc} A +\I B & C +\I D \\ A\rq{} +\I B\rq{} & C\rq{} +\I D\rq{} \end{array} \right] ,
\]
we have $A = -A\t = C\rq{}$, $B = -D\rq{} = B\t$, $C = -A\rq{} = C\t$, $D = B\rq{} = D\t$. According to the map (\ref{map}), the Lie algebra of $\h_{{\rm Sp}(d/4)}$ with $d=4n$ is
\[
	\mfh_{{\rm Sp}(d/4)} = \Big\{ \left[ \begin{array}{cccc}
		A & C & B & D \\
		-C & A & D & -B \\
		-B & -D & A & C \\
		-D & B & -C & A \\
	\end{array} \right] \, |\,
	A \in {\cal M}_- \mbox{ and } B,C,D \in {\cal M}_+ \Big\}\, .
\]
The elements of $\mfh_{{\rm Sp}(d/4)}$ can also be written as
\begin{equation}\label{gs}
	\id_2 \otimes \id_2 \otimes A +
	(\I \sigma_2) \otimes \sigma_3 \otimes B +
	\id_2 \otimes (\I \sigma_2) \otimes C +
	(\I \sigma_2) \otimes \sigma_1 \otimes D .
\end{equation}
In this form, it is easy to see that the matrices that commute with $\mfh_{{\rm Sp}(d/4)}$ are
\begin{equation}\label{comm Sp}
	z_0\, \id_{4n} +
	z_1\, \sigma_1 \otimes (\I \sigma_2) \otimes \id_n +
	z_2\, (\I \sigma_2) \otimes\id_2 \otimes \id_n +
	z_3\, \sigma_3 \otimes (\I \sigma_2) \otimes \id_n
\end{equation}
with $z_0, z_1, z_2, z_3 \in\reals$ if we want them to be real. A fact used in~(\ref{M3}) is that: all symmetric matrices of the above form are proportional to the identity. If we restrict to the antisymmetric ones ($z_0 =0$) we obtain the Lie algebra of the group ${\cal F}_{{\rm SU}(2)}$, defined in (\ref{s1}-\ref{s3}); hence $\h_{{\rm Sp}(d/4)}$ commutes with ${\cal F}_{{\rm SU}(2)}$. By setting $A=C=D=\zm$ and $B=\id_n$ in~(\ref{gs}) one can see that $X= (\I \sigma_2) \otimes\sigma_3 \otimes \id_n \in \mfh_{{\rm Sp}(d/4)}$. Since $X^2 = -\id_d$ we have $\e^{\pi X} = -\id_d \in \h_{{\rm Sp}(d/4)}$.

\subsection{Transitivity}

The transitivity of $\h_{{\rm Sp}(d/4)}$ follows from the transitivity of ${\rm Sp}(d/4)$ in the set of unit vectors in $\mathbb{H}^{d/4}$, where $\mathbb{H}$ are the quaternions~\cite{group_book}.

\subsection{Uniqueness}

In what follows we show that $\h_{{\rm Sp}(d/4)}$ is the unique $d$-dimensional real representation of ${\rm Sp}(d/4)$ which is $\reals$-irreducible.

\begin{lemma}
For $n\geq 3$, the only non-trivial $\mathbb{C}$-irreducible representation of $\mathfrak{sp}(n)$ with dimension smaller or equal than $4n$ is the fundamental one, defined in~(\ref{sp}).
\end{lemma}
\begin{proof}
There is a one-to-one correspondence between $\mathbb{C}$-irreducible representations of $\mathfrak{sp}(n)$ and $n$-dimensional vectors of natural numbers $(a_1,\ldots, a_{n})$, which are the coefficients of the dominant weight when it is expressed in the basis of the fundamental weights~\cite{group_book}. The dimension of the representation is given by
\begin{eqnarray}\label{Dsp}
	&&D(a_1,\ldots, a_n) =
\\  \nonumber
	&&\prod_{j=1}^n \prod_{i=1}^{j-1}
	\left( 1 + \frac{a_i, \ldots + a_{j-1}}{j-i}\right)
	\prod_{j=1}^n \prod_{i=1}^j
	\left( 1 + \frac{a_i+\ldots + a_{j-1}+2(a_j+\ldots + a_n)}{2 n+2-i-j}\right)\ ,
\end{eqnarray}
where the trivial representation corresponds to $D(0,\ldots,0)=1$~\cite{group_book}.

Let us first consider the case $n\geq 3$. According to~\cite{Haar-book} we have
\[
   f_j:= D(0,\ldots,0,\underbrace{1}_j,0,\ldots,0)={2n \choose j}-{2n \choose {j-2}},
\]
interpreting ${2n \choose {-1}}=0$. Thus, $D(1,0,\ldots,0)=2n$, giving the fundamental representation mentioned in the statement of the lemma.
Now let $2\leq j\leq n-1$, then
\[
   f_j = \frac{2(n-j+1)}{2n+2}{{2n+2}\choose j}\geq \frac 4 {2n+2}{{2n+2}\choose j}\geq \frac 4 {2n+2}{{2n+2}\choose 2}=4n+2>4n.
\]
Furthermore, if $n\geq 3$, then
\[
   f_n=\frac{2(2n+1)!}{n!(n+2)!} > 4n.
\]
It follows from strict monotonicity of $D$ that any further irrep of $Sp(n)$ of dimension $\leq 4n$ must have coefficients $a_2=\ldots=a_n=0$.
A quick calculation shows that $D(2,0,\ldots,0)=n(2n+1)>4n$ for $n\geq 3$, so there are no further possibilities of this kind.
\end{proof}

The case $n=2$ can be explored numerically, where strict monotonicity of $D$ makes sure that we do not overlook any possibilities. This exploration shows that the dimensions of the smallest $\mathbb{C}$-irreducible representations of $\mathfrak{sp}(2)$ are $1, 4, 5, 10, \ldots$, without repetitions, hence they are self-dual. Since $1+1< 4\cdot 2$ and $5+5 > 4\cdot 2$, the only possibility is $4+4=4\cdot 2$.

\section{Product groups}

\subsection{Description}

In the previous two sections we have described the matrix groups $\h_{{\rm SU}(d/2)}$ and $\h_{{\rm Sp}(d/4)}$, and obtained the sets of matrices which commute with them: (\ref{comm SU}) and (\ref{comm Sp}) respectively. By removing the part proportional to the identity in (\ref{comm SU}) and (\ref{comm Sp}), we obtain the Lie algebras of ${\cal F}_{{\rm SU}(d/2)}$ and ${\cal F}_{{\rm Sp}(d/4)}$ respectively. The $\h$-representation of the product groups is
\begin{eqnarray*}
	\h_{{\rm U}(d/2)} = \h_{{\rm SU}(d/2) \times {\rm U}(1)} &=& \h_{{\rm SU}(d/2)} {\cal F}_{{\rm U}(1)} \ , \\
	\h_{{\rm Sp}(d/4) \times {\rm U}(1)} &=& \h_{{\rm Sp}(d/4)} {\cal F}_{{\rm U}(1)}\ , \\
	\h_{{\rm Sp}(d/4) \times {\rm SU}(2)} &=& \h_{{\rm Sp}(d/4)} {\cal F}_{{\rm SU}(2)}\ . \\
\end{eqnarray*}

\subsection{Transitivity}

Transitivity follows from the fact that the first factors of the products, $\h_{{\rm SU}(d/2)}$ and $\h_{{\rm Sp}(d/4)}$, are transitive.

\subsection{Uniqueness}

According to~\cite{Montgomery}, if a product group is transitive on the sphere then one of its factors is either ${\rm SU}(2)$ or ${\rm U}(1)$, and the other factor is transitive on the sphere as well. We have seen above that $\h_{{\rm SU}(d/2)}$ and $\h_{{\rm Sp}(d/4)}$ are the unique representations which are transitive on the sphere. We have also seen that the Lie algebras of ${\cal F}_{{\rm U}(1)}$ and ${\cal F}_{{\rm SU}(2)}$ are the unique  ones which commute with the previous matrix groups (up to linear equivalence). This implies that the above described representations of the product groups from Table~\ref{groups trans} are the unique ones which are transitive on the sphere.

\section{The spin groups}

The $\h$-representation of ${\rm Spin}(9)$ and ${\rm Spin}(7)$ are linearly equivalent to their fundamental representations, which are described in what follows.

\subsection{Description of $\h_{{\rm Spin}(9)}$}

First, we describe a representation of the Clifford group associated to ${\rm Spin}(9)$ (more details can be found on page~68 of~\cite{Haar-book}). The generators
\[
\begin{array}{ll}
	\gamma_1 = \sigma_1 \otimes \id \otimes \id \otimes \id\ , &
	\gamma_2 = \sigma_3 \otimes \id \otimes \id \otimes \id\ ,
\\
	\gamma_3 = \sigma_2 \otimes \sigma_1 \otimes \id \otimes \id\ , &
	\gamma_4 = \sigma_2 \otimes \sigma_3 \otimes \id \otimes \id\ ,
\\
	\gamma_5 = \sigma_2 \otimes \sigma_2 \otimes \sigma_1 \otimes \id\ , &
	\gamma_6 = \sigma_2 \otimes \sigma_2 \otimes \sigma_3 \otimes \id\ ,
\\
	\gamma_7 = \sigma_2 \otimes \sigma_2 \otimes \sigma_2 \otimes \sigma_1\ , &
	\gamma_8 = \sigma_2 \otimes \sigma_2 \otimes \sigma_2 \otimes \sigma_3\ ,
\\
	\gamma_9 = \sigma_2 \otimes \sigma_2 \otimes \sigma_2 \otimes \sigma_2\ , &
\end{array}
\]
satisfy the Clifford relations
\[
	\gamma_i \gamma_j + \gamma_j \gamma_i = 2\delta_{i,j} \id\ .
\]
The Clifford group ${\rm CL}(9)$ consists of all possible products of the generators $\gamma_1, \ldots ,\gamma_9$. Some of these elements are not real---for example $\gamma_3 \notin \reals^{16\times 16}$. However, the fact that it is irreducible~\cite{group_book}
\[
	\frac{1}{|{\rm CL}(9)|} \sum_{G\in {\rm CL}(9)} \tr(G^*) \tr(G) = 1
\quad \mbox{ and } \quad
	\frac{1}{|{\rm CL}(9)|} \sum_{G\in {\rm CL}(9)} \tr(G^2) = 1\ ,
\]
implies that there is a unitary matrix $U \in \mathbb{C}^{16 \times 16}$ such that $\{U G U^\dagger |\, G\in {\rm CL}(9) \} \subseteq \reals^{16\times 16}$ is an equivalent representation of ${\rm CL}(9)$ where all matrices are real. To obtain $U$ we use Schur\rq{}s lemma~\cite{group_book}
\[
	\frac{1}{|{\rm CL}(9)|} \sum_{G\in {\rm CL}(9)}
	(U G U^\dagger)\t (U G U^\dagger)
	=\id \ ,
\]
or equivalently
\[
	\frac{1}{|{\rm CL}(9)|} \sum_{G\in {\rm CL}(9)}
	G\t (U\t U) G
	= (U\t U) \ .
\]
This gives a homogenous linear system for the unknown $(U\t U)$. The set of solutions constitutes a linear space of dimension one. Since we know that $(U\t U)$ is unitary, we choose the solution
\[
	(U\t U) =
	\left[ \begin{array}{cccccccccccccccc}
		0 & 0 & 0 & 0 & 0 & 1 & 0 & 0 & 0 & 0 & 0 & 0 & 0 & 0 & 0 & 0 \\
		0 & 0 & 0 & 0 &-1 & 0 & 0 & 0 & 0 & 0 & 0 & 0 & 0 & 0 & 0 & 0 \\
		0 & 0 & 0 & 0 & 0 & 0 & 0 & 1 & 0 & 0 & 0 & 0 & 0 & 0 & 0 & 0 \\
		0 & 0 & 0 & 0 & 0 & 0 &-1 & 0 & 0 & 0 & 0 & 0 & 0 & 0 & 0 & 0 \\
		0 &-1 & 0 & 0 & 0 & 0 & 0 & 0 & 0 & 0 & 0 & 0 & 0 & 0 & 0 & 0 \\
		1 & 0 & 0 & 0 & 0 & 0 & 0 & 0 & 0 & 0 & 0 & 0 & 0 & 0 & 0 & 0 \\
		0 & 0 & 0 &-1 & 0 & 0 & 0 & 0 & 0 & 0 & 0 & 0 & 0 & 0 & 0 & 0 \\
		0 & 0 & 1 & 0 & 0 & 0 & 0 & 0 & 0 & 0 & 0 & 0 & 0 & 0 & 0 & 0 \\
		0 & 0 & 0 & 0 & 0 & 0 & 0 & 0 & 0 & 0 & 0 & 0 & 0 & 1 & 0 & 0 \\
		0 & 0 & 0 & 0 & 0 & 0 & 0 & 0 & 0 & 0 & 0 & 0 &-1 & 0 & 0 & 0 \\
		0 & 0 & 0 & 0 & 0 & 0 & 0 & 0 & 0 & 0 & 0 & 0 & 0 & 0 & 0 & 1 \\
		0 & 0 & 0 & 0 & 0 & 0 & 0 & 0 & 0 & 0 & 0 & 0 & 0 & 0 &-1 & 0 \\
		0 & 0 & 0 & 0 & 0 & 0 & 0 & 0 & 0 &-1 & 0 & 0 & 0 & 0 & 0 & 0 \\
		0 & 0 & 0 & 0 & 0 & 0 & 0 & 0 & 1 & 0 & 0 & 0 & 0 & 0 & 0 & 0 \\
		0 & 0 & 0 & 0 & 0 & 0 & 0 & 0 & 0 & 0 & 0 &-1 & 0 & 0 & 0 & 0 \\
		0 & 0 & 0 & 0 & 0 & 0 & 0 & 0 & 0 & 0 & 1 & 0 & 0 & 0 & 0 & 0
	\end{array} \right]\ .
\]
It is straightforward to check that the unitary
\[
	U = \frac{1}{\sqrt{2}}
	\left[ \begin{array}{cccccccccccccccc}
		1 & 0 & 0 & 0 & 0 & 1 & 0 & 0 & 0 & 0 & 0 & 0 & 0 & 0 & 0 & 0 \\
		0 & \I & 0 & 0 & \I & 0 & 0 & 0 & 0 & 0 & 0 & 0 & 0 & 0 & 0 & 0 \\
		0 & 0 & 1 & 0 & 0 & 0 & 0 & 1 & 0 & 0 & 0 & 0 & 0 & 0 & 0 & 0 \\
		0 & 0 & 0 & \I & 0 & 0 & \I & 0 & 0 & 0 & 0 & 0 & 0 & 0 & 0 & 0 \\
		0 &-1 & 0 & 0 & 1 & 0 & 0 & 0 & 0 & 0 & 0 & 0 & 0 & 0 & 0 & 0 \\
		\I & 0 & 0 & 0 & 0 &-\I & 0 & 0 & 0 & 0 & 0 & 0 & 0 & 0 & 0 & 0 \\
		0 & 0 & 0 &-1 & 0 & 0 & 1 & 0 & 0 & 0 & 0 & 0 & 0 & 0 & 0 & 0 \\
		0 & 0 & \I & 0 & 0 & 0 & 0 &-\I & 0 & 0 & 0 & 0 & 0 & 0 & 0 & 0 \\
		0 & 0 & 0 & 0 & 0 & 0 & 0 & 0 & 1 & 0 & 0 & 0 & 0 & 1 & 0 & 0 \\
		0 & 0 & 0 & 0 & 0 & 0 & 0 & 0 & 0 & \I & 0 & 0 & \I & 0 & 0 & 0 \\
		0 & 0 & 0 & 0 & 0 & 0 & 0 & 0 & 0 & 0 & 1 & 0 & 0 & 0 & 0 & 1 \\
		0 & 0 & 0 & 0 & 0 & 0 & 0 & 0 & 0 & 0 & 0 & \I & 0 & 0 & \I & 0 \\
		0 & 0 & 0 & 0 & 0 & 0 & 0 & 0 & 0 &-1 & 0 & 0 & 1 & 0 & 0 & 0 \\
		0 & 0 & 0 & 0 & 0 & 0 & 0 & 0 & \I & 0 & 0 & 0 & 0 &-\I & 0 & 0 \\
		0 & 0 & 0 & 0 & 0 & 0 & 0 & 0 & 0 & 0 & 0 &-1 & 0 & 0 & 1 & 0 \\
		0 & 0 & 0 & 0 & 0 & 0 & 0 & 0 & 0 & 0 & \I & 0 & 0 & 0 & 0 &-\I
	\end{array} \right]\ ,
\]
satisfies the above equation. The Lie algebra of $\h_{{\rm Spin}(9)}$ is
\begin{equation}\label{la}
	\mfh_{{\rm Spin}(9)} = \Big\{ \mbox{$\sum_{1 \leq i < j \leq 9}$}\, c_{ij}\, U\gamma_i \gamma_j U^\dagger\, |\, c_{ij} \in \reals \Big\}\ .
\end{equation}
(Note that this is isomorphic to $\mathfrak{so}(9)$.) Since $(U \gamma_1 \gamma_2 U^\dagger)^2= -\id$ we have $\e^{\theta U \gamma_1 \gamma_2U^\dagger} = \id \cos \theta + U\gamma_1 \gamma_2 U^\dagger \sin \theta$ which is minus the identity for $\theta=\pi$, hence $-\id \in \h_{{\rm Spin}(9)}$.

\subsection{Description of $\h_{{\rm Spin}(7)}$}

Let us start by describing a representation of the even part of the Clifford group associated to ${\rm Spin}(7)$ (more details can be found in page~68 of~\cite{Haar-book}). The generators
\[
\begin{array}{ll}
	\gamma_1 = \sigma_1 \otimes \id \otimes \id\ , &
	\gamma_2 = \sigma_3 \otimes \id \otimes \id\ ,
\\
	\gamma_3 = \sigma_2 \otimes \sigma_1 \otimes \id\ , &
	\gamma_4 = \sigma_2 \otimes \sigma_3 \otimes \id\ ,
\\
	\gamma_5 = \sigma_2 \otimes \sigma_2 \otimes \sigma_1\ , &
	\gamma_6 = \sigma_2 \otimes \sigma_2 \otimes \sigma_3\ ,
\\
	\gamma_7 = \sigma_2 \otimes \sigma_2 \otimes \sigma_2\ , &
\end{array}
\]
satisfy the Clifford relations
\[
	\gamma_i \gamma_j + \gamma_j \gamma_i = 2\delta_{i,j} \id\ .
\]
The Clifford group ${\rm CL}(7)$ consists of all possible products of the generators $\gamma_1, \ldots ,\gamma_7$. Contrary to ${\rm CL}(9)$, this representation is not equivalent to a real one~\cite{group_book}, since
\[
	\frac{1}{|{\rm CL}(7)|} \sum_{G\in {\rm CL}(7)} \tr(G^*) \tr(G) = 1\
	\mbox{ and }\
	\frac{1}{|{\rm CL}(7)|} \sum_{G\in {\rm CL}(7)} \tr(G^2) = 0\ .
\]
However, in order to construct the Lie algebra of ${\rm Spin}(7)$ ---as in~(\ref{la})--- we only need a real representation for the even subgroup ${\rm CL}_+ (7) \leq {\rm CL}(7)$. The even subgroup ${\rm CL}_+(7)$ consists of all possible products of an even number of generators $\gamma_1, \ldots ,\gamma_7$. This is irreducible
\[
	\frac{1}{|{\rm CL}_+ (7)|} \sum_{G\in {\rm CL}_+ (7)} \tr(G^*) \tr(G) = 1\ ,
\]
and equivalent to a real representation
\[
	\frac{1}{|{\rm CL}_+ (7)|} \sum_{G\in {\rm CL}_+ (7)} \tr(G^2) = 1\ .
\]
Therefore, there is a unitary matrix $U \in \mathbb{C}^{8 \times 8}$ such that $\{U G U^\dagger : \forall G\in {\rm CL}_+ (7) \} \subseteq \reals^{8\times 8}$ is an equivalent representation of ${\rm CL}_+ (7)$ where all matrices are real. To obtain $U$ we use Schur\rq{}s lemma~\cite{group_book}
\[
	\frac{1}{|{\rm CL}_+ (7)|} \sum_{G\in {\rm CL}_+ (7)}
	G\t (U\t U) G
	= (U\t U) \ .
\]
This gives a homogenous linear system for the unknown $(U\t U)$. The set of solutions constitutes a linear space of dimension one. Since we know that $(U\t U)$ is unitary, we chose the solution
\[
	(U\t U) =
	\left[ \begin{array}{cccccccc}
		0 & 0 & 0 & 0 & 0 & 1 & 0 & 0 \\
		0 & 0 & 0 & 0 &-1 & 0 & 0 & 0 \\
		0 & 0 & 0 & 0 & 0 & 0 & 0 & 1 \\
		0 & 0 & 0 & 0 & 0 & 0 &-1 & 0 \\
		0 &-1 & 0 & 0 & 0 & 0 & 0 & 0 \\
		1 & 0 & 0 & 0 & 0 & 0 & 0 & 0 \\
		0 & 0 & 0 &-1 & 0 & 0 & 0 & 0 \\
		0 & 0 & 1 & 0 & 0 & 0 & 0 & 0 \\
	\end{array} \right]\ .
\]
It is straightforward to check that the unitary
\[
	U = \frac{1}{\sqrt{2}}
	\left[ \begin{array}{cccccccc}
		1 & 0 & 0 & 0 & 0 & 1 & 0 & 0 \\
		0 & \I & 0 & 0 & \I & 0 & 0 & 0 \\
		0 & 0 & 1 & 0 & 0 & 0 & 0 & 1 \\
		0 & 0 & 0 & \I & 0 & 0 & \I & 0 \\
		0 &-1 & 0 & 0 & 1 & 0 & 0 & 0 \\
		\I & 0 & 0 & 0 & 0 &-\I & 0 & 0 \\
		0 & 0 & 0 &-1 & 0 & 0 & 1 & 0 \\
		0 & 0 & \I & 0 & 0 & 0 & 0 &-\I \\
	\end{array} \right]\ ,
\]
satisfies the above equation. The Lie algebra of $\h_{{\rm Spin}(7)}$ is
\begin{equation}\label{la}
	\mfh_{{\rm Spin}(7)}= \Big\{ \mbox{$\sum_{1 \leq i < j \leq 7}$}\, c_{ij}\, U\gamma_i \gamma_j U^\dagger\, |\, c_{ij} \in \reals \Big\}\ .
\end{equation}
Since $(U \gamma_1 \gamma_2 U^\dagger)^2= -\id$ we have $\e^{\theta U \gamma_1 \gamma_2U^\dagger} = \id \cos \theta + U\gamma_1 \gamma_2 U^\dagger \sin \theta$ which is minus the identity for $\theta=\pi$, hence $-\id \in \h_{{\rm Spin}(7)}$.

\subsection{Transitivity}

In what follows we show that $\h_{{\rm Spin}(n)}$ is transitive in the unit sphere of $\reals^d$ where $d=2^{(n-1)/2}$ and $n=7,9$. To do this, we just have to check that $\h_{{\rm Spin}(n)}$ maps any point $\bfu$ within the unit sphere to any of its neighboring points $\bfu + \epsilon\bfv$ within the unit sphere (so $\bfv\cdot \bfu =0$); by composing infinitesimal transformations of this sort we can map any point to any other. This is shown in Lemma 1.3 from~\cite{lemma13}.

In the Lie algebra language: for any pair of orthogonal vectors $\bfu, \bfv$ there is $X \in \mfh_{{\rm Spin}(n)}$ such that $\bfv =X\bfu$, therefore $\e^{\epsilon X}\bfu =\bfu + \epsilon\bfv + {\cal O}(\epsilon^2)$. In order to show this we assume the opposite: the vector space $\mfh_{{\rm Spin}(n)}\bfu$ does not contain $\bfv$. Since $\mfh_{{\rm Spin}(n)}$ is made out of antisymmetric matrices, the vector space $\mfh_{{\rm Spin}(n)}\bfu$ is orthogonal to $\bfu$. This implies that there is ${\bf w}$ orthogonal to $\bfu$ such that
\begin{equation}
	{\bf w} \cdot U\gamma_i \gamma_j U^\dagger \bfu =0
	\quad\mbox{ for all } 0 \leq i < j \leq n\ .
\end{equation}
Therefore the vectors $\gamma_1 U^\dagger \bfu, \ldots, \gamma_n U^\dagger \bfu, \gamma_1 U^\dagger {\bf w}, \ldots, \gamma_n U^\dagger {\bf w}$ are orthogonal, but there cannot be $2n$ orthogonal vectors in a $2^{(n-1)/2}$-dimensional space when $n=7,9$, hence the contradiction. Thus, the group is transitive.

Note that this argumentation does not apply to $n=3, 5$, since the corresponding groups ${\rm Spin}(n)$ do not have real, $2^{(n-1)/2}$-dimensional, $\mathbb{C}$-irreducible representations.

\subsection{Uniqueness}

According to~\cite{group_book},  there is only one nontrivial $\mathbb{C}$-irreducible representation of ${\rm Spin}(n)$ of dimension not larger than $d=2^{(n-1)/2}$, for odd $n$.

\section{$\mathrm{G}_2$}
\label{SecG2}

\subsection{Description}

The $\h$-representation of ${\rm G}_2$ is its fundamental one, hence $\mfh_{{\rm G}_2} = \mathfrak{g}_2$. In page~33 from~\cite{Arenas} one can find the following parametrization of $\mathfrak{g}_2$:
\begin{equation}\label{generic G2}
	\left[
      \begin{array}{ccccccc}
         0 & r_3 & -r_2 & r_5 & -r_4 & -r_7 & -r_6+s_6 \\
         -r_3 & 0 & r_1 & r_6 & -r_7 +s_7 & r_4-s_4 & r_5+s_5 \\
         r_2 & -r_1 & 0 & -s_7 & s_6 & s_5 & s_4 \\
         -r_5 & -r_6 & s_7 & 0 & -r_1+s_1 & -r_2+s_2 & -r_3+s_3 \\
         r_4 & r_7-s_7 & -s_6 & r_1-s_1 & 0 & s_3 & -s_2 \\
         r_7 & -r_4+s_4 & -s_5 & r_2-s_2 & -s_3 & 0 & s_1 \\
         r_6-s_6 & -r_5-s_5 & -s_4 & r_3-s_3 & s_2 & -s_1 & 0
      \end{array}
   \right] \in \mathfrak{g}_2\ ,
\end{equation}
for all $r_1,\ldots, r_7, s_1,\ldots, s_7 \in\reals$. The stabilizer subgroup on $\bfe_1$ is $\h_1 = \{G\in \h_{{\rm G}_2} \,|\, G\bfe_1 = \bfe_1\}$, and the corresponding Lie algebra is $\mfh_1 = \{X\in \mathfrak{g}_2 \,|\, X\bfe_1 = \zm\}$. This is the set of matrices like~(\ref{generic G2}) with $r_2= r_3= r_4= r_5= r_7=0$ and $r_6 = s_6$. Solving a system of linear equations one can see that any antisymmetric $7\times 7$ real matrix which commutes with $\mfh_1$ is proportional to
\begin{equation}\label{def T}
   T=\left[
      \begin{array}{ccccccc}
         0 & 0 & 0 & 0 & 0 & 0 & 0 \\
         0 & 0 & 1 & 0 & 0 & 0 & 0 \\
         0 & -1 & 0 & 0 & 0 & 0 & 0 \\
         0 & 0 & 0 & 0 & 1 & 0 & 0 \\
         0 & 0 & 0 & -1 & 0 & 0 & 0 \\
         0 & 0 & 0 & 0 & 0 & 0 & -1 \\
         0 & 0 & 0 & 0 & 0 & 1 & 0
      \end{array}
   \right].
\end{equation}
Note that $T$ is not an element of $\mathfrak{g}_2$ (this is relevant in Section~\ref{G2s}).

Setting $r_3=1$ and all the rest of parameters equal to zero in~(\ref{generic G2}) we obtain
\[
   X=\left[
      \begin{array}{ccccccc}
         0 & 1 & 0 & 0 & 0 & 0 & 0 \\
         -1 & 0 & 0 & 0 & 0 & 0 & 0 \\
         0 & 0 & 0 & 0 & 0 & 0 & 0 \\
         0 & 0 & 0 & 0 & 0 & 0 & -1 \\
         0 & 0 & 0 & 0 & 0 & 0 & 0 \\
         0 & 0 & 0 & 0 & 0 & 0 & 0 \\
         0 & 0 & 0 & 1 & 0 & 0 & 0
      \end{array}
   \right] \in \mathfrak{g}_2\ .
\]
By exponentiating this we obtain an element of the group which plays a special role in Section~\ref{G2s}
\begin{equation}\label{def H}
	H= \e^{\pi X}=
	\left[
      \begin{array}{ccccccc}
         -1& 0 & 0 & 0 & 0 & 0 & 0 \\
         0 &-1 & 0 & 0 & 0 & 0 & 0 \\
         0 & 0 & 1 & 0 & 0 & 0 & 0 \\
         0 & 0 & 0 &-1 & 0 & 0 & 0 \\
         0 & 0 & 0 & 0 & 1 & 0 & 0 \\
         0 & 0 & 0 & 0 & 0 & 1 & 0 \\
         0 & 0 & 0 & 0 & 0 & 0 &-1
      \end{array}
   \right] \in \h_{{\rm G}_2}\ .
\end{equation}

\subsection{Transitivity and uniqueness}

Transitivity is shown in page~364 from~\cite{group_book}. Uniqueness follows from the fact that the fundamental representation of ${\rm G}_2$ is the unique non-trivial $\mathbb{C}$-irreducible representation with dimension equal or less than 7.



\end{document}